\documentclass[sn-mathphys-num]{sn-jnl}

\usepackage{graphicx}%
\usepackage{multirow}%
\usepackage{amsmath,amssymb,amsfonts}%
\usepackage{amsthm}%
\usepackage{mathrsfs}%
\usepackage[title]{appendix}%
\usepackage{xcolor}%
\usepackage{textcomp}%
\usepackage{manyfoot}%
\usepackage{booktabs}%
\usepackage{algorithm}%
\usepackage{algorithmicx}%
\usepackage{algpseudocode}%
\usepackage{listings}%

\usepackage[english]{babel}
\usepackage{amssymb}
\usepackage{underscore}
\usepackage{empheq}

\usepackage{float}
\usepackage{amsmath}
\usepackage{mathtools}
\usepackage[super]{nth}
\usepackage{caption}
\usepackage{subcaption}
\usepackage{verbatim} 

\usepackage{etoolbox}
\DeclareMathAlphabet{\mathpzc}{OT1}{pzc}{m}{it}
\usepackage{tabularx}
\usepackage{lipsum}
\usepackage{multicol}
\usepackage{afterpage}
\usepackage{setspace}
\usepackage{parskip}

\newcommand{\DS}{\mathrm{\Omega}}             
\newcommand{\DT}{{I}}                  
\newcommand{\vx}{\vect{x}} 
\newcommand{\PGDm}{m} 

\newcommand{\PGDmz}{m_{z}} 

\newcommand{\SDvect}[1]{%
            \underline{\boldsymbol{\mathbf{%
                \mathit{#1}
            }}}
}

\newcommand{\vect}[1]{\underline{{#1}}} 

\newcommand{\ctens}[1]{\underline{\underline{{#1}}}} 

\newcommand{\dotp}{\cdot} 

\newcommand{\pgdu}{\bar{u}} 

\newcommand{\Su}{\mathcal{U}} %
\newcommand{\Esig}{\mathcal{F}} 

\newcommand{\eps}{\varepsilon}

\newcommand{\R}{\mathbb{R}} 

\newcommand{\normi}[1]{\left\|{#1}\right\|}

\newcommand{\pgdt}{\lambda}

\theoremstyle{thmstyleone}%
\theoremstyle{thmstyletwo}%

\theoremstyle{thmstylethree}%

\begin{document}

\title[Article Title]{A reduced simulation applied to viscoelastic fatigue
of polymers using a time multi-scale approach based on Partition of Unity method}


\author*[1,5]{\fnm{Sebastian} \sur{Rodriguez}}\email{sebastian.rodriguez\_iturra@ensam.eu}
\equalcont{These authors contributed equally to this work.}

\author[1,5]{\fnm{Angelo} \sur{Pasquale}}\email{angelo.pasquale@ensam.eu}
\equalcont{These authors contributed equally to this work.}

\author[1]{\fnm{Jad} \sur{Mounayer}}\email{jad.mounayer@ensam.eu}
\equalcont{These authors contributed equally to this work.}

\author[2]{\fnm{Diego} \sur{Canales}}\email{dcanales@uloyola.es}
\equalcont{These authors contributed equally to this work.}

\author[3]{\fnm{Marianne} \sur{Beringhier}}\email{marianne.beringhier@ensma.fr}
\equalcont{These authors contributed equally to this work.}

\author[1]{\fnm{Chady} \sur{Ghnatios}}\email{chady.ghnatios@ensam.eu}
\equalcont{These authors contributed equally to this work.}

\author[4]{\fnm{Amine} \sur{Ammar}}\email{amine.ammar@ensam.eu}
\equalcont{These authors contributed equally to this work.}

\author[1,5,6]{\fnm{Francisco} \sur{Chinesta}}\email{francisco.chinesta@ensam.eu}
\equalcont{These authors contributed equally to this work.}

\affil*[1]{\orgname{PIMM Lab, Arts et M\'etiers Institute of Technology}, \orgaddress{\street{151 Boulevard de l'H\^opital}, \city{Paris}, \postcode{75013}, \country{France}}}

\affil[2]{\orgname{Departamento de M\'etodos Cuantitativos, Universidad Loyola Andaluc\'ia}, \orgaddress{\street{Av. de las Universidades, 2, Dos Hermanas}, \city{Sevilla}, \postcode{41704}, \country{Spain}}}

\affil[3]{\orgname{Institut P', D\'epartement Physique et m\'ecanique des mat\'eriaux, UPR CNRS 3346, ISAE - ENSMA}, \orgaddress{ \city{Chasseneuil-du-Poitou}, \postcode{86360}, \country{France}}}

\affil[4]{\orgname{LAMPA Lab, Arts et M\'etiers Institute of Technology}, \orgaddress{\street{2, Boulevard du Ronceray, BP 93525}, \city{Angers}, \postcode{49035}, \country{France}}}

\affil[5]{\orgname{ESI Group Chair @ Arts et M\'etiers Institute of Technology}, \orgaddress{\street{151 Boulevard de l'H\^opital}, \city{Paris}, \postcode{75013}, \country{France}}}

\affil[6]{\orgname{CNRS@CREATE LTD}, \orgaddress{\street{1 Create Way, \#08-01 CREATE Tower}, \postcode{138602}, \country{Singapore}}}


\abstract{The simulation of 
viscoelastic time-evolution problems described by a large number of internal variables and with a
large spectrum of relaxation times requires high computational resources for their resolution. Furthermore, the internal variables evolution is 
described by a set of linear differential equations which involves many
time scales. In this context, the use of a space-time PGD approximation is proposed here to boost their resolution, where the temporal functions are constructed following a multi-scale strategy along with the Partition of Unity method, in order to catch each dynamic efficiently. The feasibility and the robustness of the method are discussed in the case of a polymer in a non-equilibrium state under cyclic loading.}




\maketitle

\section{Introduction}

The Proper Generalized Decomposition (PGD) \cite{ammar2006new} is a numerical method for approximating the solutions of multidimensional Partial Differential Equations (PDEs). The PGD enables the construction of a reduced model of a problem beforehand, integrating its approximation directly into the PDEs while solving the problem. This process is iterative, relying on a minimization problem. It has been extensively applied in various domains, including stochastic frameworks and multidimensional scenarios, showcasing its versatility and efficacy \cite{ladeveze2003multiscale, nouy2010priori, nouy2007generalized, ammar2007new, chinesta2010proper, chinesta2011short, pruliere2010deterministic, chinesta2013pgd}.

In \cite{hammoud2011application,hammoud2014reduced} the PGD has been applied to predict viscoelastic polymers' behavior in a non-equilibrium state under creep and cyclic loading. In these works, local differential equations describing the internal variables evolution (describing the dissipative phenomena) are strongly coupled with a global equilibrium equation. The PGD-based space-time separation is applied considering a globalization of the local equations and a fixed point algorithm between the displacement field and the internal variables. 

In particular, in \cite{hammoud2014reduced}, up to 50 internal variables have been considered, showing the potential of the PGD in solving real viscoelastic problems under creep and cyclic loading. A special focus of \cite{hammoud2014reduced} has been the link between relaxation times and time discretization adopted within the numerical method when the sought solution evolves at different time scales (the cycle time and the total time). Indeed, in the presence of many time scales, one linked to mechanical loading and the others linked to the relaxation times of the internal variables, the solution may be highly expensive computationally. 

As pointed out in \cite{hammoud2014reduced}, the limits of the procedure stand in the computational cost of cyclic fatigue scenarios. Indeed, in polymer materials, understanding cyclic behavior is even more challenging than in metals since they do not quickly stabilize after a few fatigue cycles. Instead, cycling evolves slowly due to creep at average stress, influenced by temperature coupling, especially at high solicitation frequencies \cite{berrehili2010multiaxial}. This acerbates the numerical simulation, as each cycle must be simulated with an appropriate time step until the last cycle.

In recent years, many works have been conducted to efficiently account for different time scales within the PGD framework, with a particular emphasis on cyclic fatigue scenarios \cite{ammar2011,badias2017,ibanez2019multiscale,pasquale2021separated,PASQUALE202375,rodriguez2024time}.

In \cite{ammar2011}, authors have developed a technique based on separation of variables, which is straightforwardly introduced in the PGD procedure. 
The time variable $t$ is expressed as two separated coordinates, via a macrotime $T$ spanning partitioned coarse times and microtime $\tau$ resolving fast responses through subdomain discretization. However, to ensure the continuity in the resulting two-scale discretization, Lagrange multipliers were employed, substantially complicating the computational implementation of the procedure.

In \cite{badias2017,ibanez2019multiscale}, various PGDs are built over different subdomains and then combined using the Partition of Unity (PU) principle \cite{melenk1996partition,babuvska1997partition}. Macro shape functions satisfying the PU enable smooth transitions between different PGDs across intervals, ensuring perfect continuity. The overlap between PGD solutions in overlapping subdomains maintains continuity when multiplied by macroscopic shape functions, leveraging PU features. While effective, the computational implementation is hindered by the need to combine microscopic discretization with macroscopic PU enrichment.

In \cite{pasquale2021separated}, authors have proposed a generalization of the multi-time PGD \cite{ammar2011}, but directly relying on the discrete tensor formulation of the separated representation involved in the PGD constructor, making use of a tensor formalism. This ensured continuity in a direct manner without resorting to the use of Lagrange multipliers, penalty or the PU paradigm. The method has been successfully employed to solve multi-scale thermal and elastodynamic problems.

In \cite{PASQUALE202375}, the multi-time PGD has been successfully applied to solve history-dependent nonlinear elastoplastic problems under cyclic loading. A generic function of time $\phi(t)$ is expressed in terms of micro-macro time submodes as $\phi(t) = \sum_{j} \phi^\tau_j(\tau)\phi^T_j(T)$, where $j$ is spanning the modes, $\phi^\tau_j$ are the functions of microtime and $\phi^T_j$ the functions of macrotime. As demonstrated in \cite{PASQUALE202375}, the microtime functions may exhibit a complex highly nonlinear behavior due to the plastic deformation along the cycle, while the macrotime ones have a smooth evolution due to the slow variation across cycles.

To extend the procedure in fatigue problems including a high number of cycles, the multi-scale behavior can be exploited to further reduce computational costs. Indeed, in the recent work \cite{rodriguez2024time}, authors build a machine learning framework based on a macrotime predictor-corrector algorithm enabling a lowcost forecasting of the nonlinear elastoplastic behavior. 

The motivation beyond the current paper is to introduce the aforementioned advances in the multi-time PGD framework, to efficient deal with internal variables in cyclic viscoelasticity of polymers. To this purpose, the space-time PGD solution proposed in \cite{hammoud2014reduced} is here recast in a space-multi-time framework. As a multi-scale procedure, the Partition of Unity method as presented in \cite{ibanez2019multiscale} is exploited. The main advantage of this method consists in relaxing the continuity requirements for the temporal reconstruction. In the context of the PU method only the macro functions should be continuous, the micro ones can have any shape.

The paper is
organized as follows. The simplified (one-dimensional) viscoelastic
model is described in section \ref{sec:viscel_problem}. Section \ref{sec:PGD_resol} presents the use of PGD to solve the considered viscoelastic model, furthermore, the constrution of the temporal functions of this low-rank decomposition using a multi-scale approach and the Partition of Unity method is also presented. Numerical results addressing cyclic loading are presented in section \ref{sec:numerical_example}, where the multi-scale approximation of the PGD is illustrated. Finally, section \ref{sec:concl_persp} corresponds to the conclusions and perspectives of the present work.

\section{Viscoelastic model}\label{sec:viscel_problem}

\begin{figure}[H] 
\centering
\includegraphics[width=0.4\textwidth]{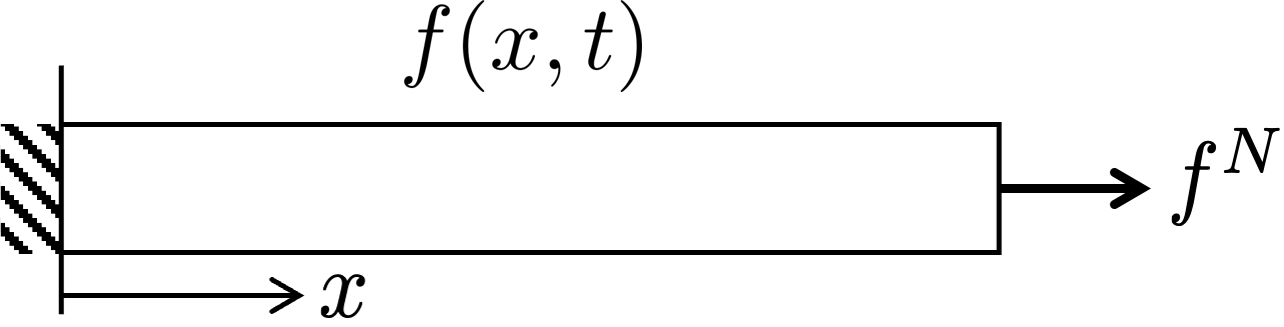}
\caption{Reference problem.}
\label{fig:ref_problem}
\end{figure}

Let us consider the one-dimensional structure of figure \ref{fig:ref_problem} occupying the spatial domain $\DS \in \R^1$, on a time domain $\DT=[0,T]$ and with constant boundary $\partial \DS = \partial_{N}\DS \oplus \partial_{D}\DS $ over time, where $\partial_{N}\DS$ and $\partial_{D}\DS$ are the boundaries related to the imposed Neumann and Dirichlet conditions respectively.
This structure is submitted to surface forces $f^{N}$ on $ \partial_{N}\DS \times \DT$ (Neumann boundary conditions), to imposed displacements $u^{D}$ on $\partial_{D}\DS \times \DT$ (Dirichlet boundary condition) and to volumetric forces $f$ on $\DS \times \DT$. 
The reference problem consists in finding a displacement field $u(x,t) \in \Su$ and a stress field $\sigma(x,t) \in \Esig$ verifying:

\begin{enumerate}
  \item Initial conditions:

$\text{on} \ \DS$,
  \begin{equation}
    \label{EQ:REF:CLim}
     \begin{array}{c} u |_{\substack{t=0}}= 0 
     \end{array}     
  \end{equation}

\item Static equilibrium equation:

$\text{on} \ \DS\times\DT$,
  \begin{equation}\label{eq.1}
  \frac{\partial {\sigma}}{\partial{x}} + f = 0, 
  \end{equation}
with $f(x,t)$ the volumetric load which depends on space and
time.
  
\item Neumann boundary conditions:

$\text{on} \ \partial_{N}\DS \times \DT$,
\begin{equation}
\label{EQ:REF:Neumann}  
    \sigma \dotp n = f^{N}    
\end{equation}
with $n$ the normal vector to the surface of $\partial_{N}\DS$. 

\item Dirichlet boundary conditions:

$\text{on} \ \partial_{D}\DS \times \DT$,
\begin{equation}\label{EQ:REF:Dirichlet} u = u^{D} = 0
 \end{equation}
 
\item Internal variables and constitutive relations:

A particular viscoelastic behavior described by internal variables $z^{[j]}$ is considered in this paper. These internal variables are determined by solving:
\begin{equation}\label{eq.2} 
\frac{dz^{[j]}}{dt} + \frac {1}{\tau_{j}} \left(z^{[j]}-z_{\infty}^{[j]}
\right) = 0 \,\,\,\,\,\,\,\,\,\,\,\, \,\,\,\, 1 \leq j \leq N 
\end{equation}
where:
\begin{equation} \label{eq.4}
z_{\infty}^{[j]} = E_{\infty}^{rj} \frac{\partial {u}}{\partial{x}}
\,\,\,\,\,\,\,\,\,\,\,\, \,\,\,\, 1 \leq j \leq N    
\end{equation}
In turn, the evolution of the internal variables affects the stress as follows:
\begin{equation}\label{eq.3}
\sigma = E_v \frac{\partial {u}}{\partial{x}} - \sum^{N}_{j=1} z^{[j]}
\end{equation}
with $E_v$ the vitreous
modulus.

Equation \eqref{eq.2} represents the kinetic of return to
equilibrium and specifies the dependence of the  relaxation times
$\tau_{j}$ on the internal variables $z^{[j]}$. The equilibrium of process $j$ is reached when the value of the corresponding internal
variable $z^{[j]}$ is equal to its value at the equilibrium noted
$z_{\infty}^{[j]}$. This internal equilibrium depends here linearly on the
macroscopic variable $\frac{\partial{u}}{\partial{x}}$ such as
formulated in Equation \eqref{eq.4}, where the relaxed modulus at
equilibrium $E_{\infty}^{rj}$ generated by the process $j$ follows
this equation: 
\begin{equation}
\label{eq.5}
E_{\infty}^{rj} = p_j E_r \,\,\,\, , \ \forall \,1 \leq j \leq N 
\end{equation}
where $E_{r}$ represents the relaxed modulus and $p_j$ the weights given by a distribution that some authors
\cite{cunat2001dnlr} link to jump atomic fluctuations in the polymer. Three
parameters are required to define the spectrum of the distribution of
the weights: the number of decades of the spectrum range, the number of
processes and the largest relaxation time \cite{andre2003rheological}. For instance, in \cite{andre2003rheological}, authors depict the spectrum of the distribution obtained with $50$ times distributed along six decades of the time scale.

\textbf{Remark}: $E_v$ is usually measured in an
high velocity experiment compared to the smallest relaxation time and
$E_r$ in a very slow experiment. 

\end{enumerate}

This mechanical problem leads to strongly coupled linear equations
between the displacement field (global model) and the internal variables
(many local models). Indeed, the displacement influences the evolution
of the internal variables (linearly in this simple viscoelastic model)
and viceversa. Moreover, each internal variable has a specific time
scale. Thus, a large number of relaxation times must be considered
simultaneously.   

\textbf{Remark}: The subscript $j$ concerns the internal variables and it varies from $1$ to $N$, it means that the Equation \eqref{eq.2} is reported
$N$ times. The specificity of each equation is related to the relaxation time of this internal variable.

\section{PGD model-reduction method applied to the viscoelastic problem}\label{sec:PGD_resol}


To solve this problem using the PGD model-order reduction method, we first globalize the local models as suggested
in \cite{chinesta2010proper}. In this sense, the low-rank approximated solutions of $u$ and $\lbrace z^{[j]} \rbrace_{j=1}^{N}$ of the coupled problem are sought under the form:
\begin{equation}\label{eq:PGD_displacement}
u(x,t) \approx u_{\PGDm}(x,t) = \sum\limits_{i=1}^{\PGDm} \pgdu_{i}(x) \pgdt_{i}(t)
\end{equation}
\begin{equation}\label{eq:PGD_int_var}
z^{[j]}(x,t) \approx z^{[j]}_{\PGDmz^{[j]}}(x,t) = \sum\limits_{i=1}^{\PGDmz^{[j]}} \Bar{z}^{[j]}_{i}(x) \phi^{[j]}_{i}(t)
\end{equation}
with $\PGDm$ and $\PGDmz^{[j]}$ the corresponding modes of the decomposition related to displacement and internal variable $j$. 

As the displacement and the internal variables are strongly coupled, all
the unknowns could be computed at each enrichment step as in the case of thermoviscoelasticity \cite{beringhier2010solution}. Here, a fixed-point iterative procedure is chosen between the displacement and the internal variables, where $(i)$ first the displacement is computed assuming the internal variables known and $(ii)$ the internal variables are computed assuming the displacement known. The low-rank construction of the displacement as well as the internal variables is constructed incrementally, that is, one mode at a time. Their determination is presented in the following sections.

\subsection{Computation of the low-rank approximation of displacement} 

In order to use the PGD into the equations to boost the resolution, one needs the weak form related to Equation \eqref{eq.1}  \cite{zienkiewicz2000finite,zienkiewicz2005finite}:
\begin{equation}\label{eq.23}
\int\limits_{\DS \times \DT}\frac{\partial u^*}{\partial x}\sigma dxdt = \int\limits_{\DS \times \DT}f u^*dxdt+\int\limits_{\partial_{N}\DS \times \DT}f^{N}u^*dxdt
\end{equation}
for all test functions $u^{*}$ selected in an appropriate functional space. 

With the stress $\sigma$ being derived from Equation \eqref{eq.3}, one can rewrite \eqref{eq.23} as follows: 
\begin{equation}\label{eq:eq_solv_disp}
\int\limits_{\DS \times \DT}\frac{\partial u^*}{\partial x} E_v \frac{\partial {u}}{\partial{x}} dxdt = \int\limits_{\DS \times \DT}f u^*dxdt + \int\limits_{\partial_{N}\DS \times \DT}f^{N} u^* dxdt + \int\limits_{\DS \times \DT} \frac{\partial u^*}{\partial x} \left( \sum^{N}_{j=1} z^{[j]} \right) dx dt
\end{equation}
%
%




%
%
%
%
Now, let's assume we have computed ``$\PGDm-1$" PGD modes such as:
%
%
\begin{equation}
u_{\PGDm}(x,t) = u_{\PGDm-1}(x,t) + \pgdu(x) \pgdt(t)
\end{equation}
therefore, the problem to be solved can be written as follows:
\begin{equation}\label{eq:eq_solv}
\begin{split}
\int\limits_{\DS \times \DT}\frac{\partial u^*}{\partial x} E_v \frac{\partial {\pgdu(x) \pgdt(t)}}{\partial{x}} dxdt = \int\limits_{\DS \times \DT}f u^*dxdt + \int\limits_{\partial_{N}\DS \times \DT}f^{N} u^* dxdt \\ + \int\limits_{\DS \times \DT} \frac{\partial u^*}{\partial x} \left( \sum^{N}_{j=1} z^{[j]} - E_v \frac{\partial {  u_{\PGDm-1}(x,t) }}{\partial{x}} \right) dx dt
\end{split}
\end{equation}
with the test function chosen as in a classical Galerkin approach \cite{chinesta2014separated}:
\begin{equation*}
u^* = \pgdu^*(x) \pgdt(t) + \pgdu(x) \pgdt^*(t)   
\end{equation*}
Equation \eqref{eq:eq_solv} is solved following a fixed-point iterative scheme to compute the spatial and temporal PGD functions until the modes stagnates.

\subsubsection{Convergence criteria for the construction of the low-rank decomposition}\label{sec:error_crit_PGD_u}

Here, the PGD decomposition of the displacement is computed until the following criteria is achieved:
\begin{equation}
\epsilon_{u} = 100 \frac{\normi{u_{\PGDm+1} - u_{\PGDm} }_{\DS \times \DT}}{\normi{ u_{\PGDm} }_{\DS \times \DT}} \leq 2 [\%]
\end{equation}
with:
\begin{equation}
\normi{\bullet}_{\DS \times \DT}^{2} = \int\limits_{\DS \times \DT} \left( \bullet \right)^{T} \left( \bullet \right) dx dt   
\end{equation}

\subsection{Computation of the low-rank approximation of internal variables} 

Once, the equilibrium problem solved and the value of the internal variables at the equilibrium
$z_{\infty}^{[j]}$ being derived from Equation \eqref{eq.4} with the value of $u_{\PGDm}$ and let's assume we have computed ``$\PGDmz^{[j]}-1$" PGD modes for the internal variable $j$ such as:

$\forall j \in [1, ..., N]$,
\begin{equation}\label{eq.16}
z^{[j]}(x,t) \approx z^{[j]}_{\PGDmz^{[j]}}(x,t) = z^{[j]}_{\PGDmz^{[j]}-1}(x,t) + \Bar{z}^{[j]}(x) \phi^{[j]}(t)
\end{equation}
In this sense, for each value of $j \in [1,N]$, one seeks to compute the $\PGDm$-mode such as it minimizes the following norm defined after \eqref{eq.2} as follows: 
\begin{equation}\label{eq:int_var_min}
\lbrace \Bar{z}^{[j]}(x),\phi^{[j]}(t) \rbrace = \underset{\lbrace \Bar{z}^{[j]}(x),\phi^{[j]}(t) \rbrace}{\text{arg min}} \normi{  \Bar{z}^{[j]}(x) \frac{\partial \phi^{[j]}(t) }{\partial t} + \frac
  {1}{\tau_{j}} \left( \Bar{z}^{[j]}(x) \phi^{[j]}(t) -z_{\infty}^{[j]} \right) + f_{res}(x,t) }_{\DS \times \DT}^{2}    
\end{equation}
with:
\begin{equation}
f_{res}(x,t) =  \frac{\partial z_{\PGDmz^{[j]}-1}^{[j]}(x,t) }{\partial t} + \frac
  {1}{\tau_{j}}  z_{\PGDmz^{[j]}-1}^{[j]}(x,t) 
\end{equation}
Equation \eqref{eq:int_var_min} is a nonlinear problem with respect to
$\Bar{z}^{[j]}(x)$ and $\phi^{[j]}(t)$. An alternating directions point fixed algorithm
is used as previously for the displacement. This iterative procedure continues until the mode stagnates.

\subsubsection{Convergence criteria for the construction of the low-rank decomposition}\label{sec:error_crit_PGD_iv}

Here, the PGD decomposition for each internal variable $j$ considered is computed until the following criteria is achieved:
\begin{equation}
\epsilon_{z}^{[j]} =  \frac{ \normi{ z^{[j]}_{\PGDmz^{[j]}} -z^{[j]}_{\PGDmz^{[j]}-1} }_{\DS \times \DT} }{ \normi{z^{[j]}_{\PGDmz^{[j]}-1}}_{\DS \times \DT}}   \leq  2 [\%]
\end{equation}

\subsection{Determination of temporal PGD functions as a multi-scale approximation throw Partition of Unity method}

Let us assume that a given function $\pgdt(t)$, is the solution of a given partial differential equation. In terms of standard approximation basis, such as finite elements, it could be expressed as:
\begin{equation}
\pgdt(t) = \sum_{i=1}^{n} N_{i}(t)q_{i}
\end{equation}
where $n$ stands for the number of dofs used in the approximation of $\pgdt(t)$, $N_{i}(t)$ for the standard finite element shape functions and $q_{i}$ the nodal value of the sought function. However, if the seek function $\pgdt(t)$ is defined on a large temporal domain the mesh has to capture the details of the solution at the finest scale, thus deriving into a prohibitive simulation cost.

In this work, since the external load considered corresponds to a fatigue excitation, a multi-scale approximation can be introduced. This is done here within the Partition of Unity
paradigm. The main idea is to enrich a coarse finite element approximation by enriching it using micro functions. Both variables that must be computed online during the resolution of the solver. In this context, we can define the following approximation:
\begin{equation}\label{eq:ms_approx}
\pgdt(t) \approx \sum_{k=1}^{\PGDm_{s}} \sum_{i=1}^{n} N_{i}(t)q_{i}^{k} \SDvect{G}^{T}(\tau(t-t_{i})) \SDvect{g}^{k}
\end{equation}
Following the temporal multi-scale PGD rationale presented in \cite{PASQUALE202375,rodriguez2024time}, the temporal PGD function $\pgdt(t)$ is approximated as the sum of $\PGDm_{s}$ sub-modes. Where $t_{i}$ is the centroid of the shape function $N_{i}(t)$, $\tau(t-t_{i})$ is a dependent variable that presents an offset based on $t_{i}$, $\SDvect{G}(\tau)$ and $\SDvect{g}^{k}$ are the vectors that contain the microscale shape function and the associated vector of DOFs respectively at sub-mode $k$.  

\begin{figure}[H] 
\centering
\includegraphics[width=0.5\textwidth]{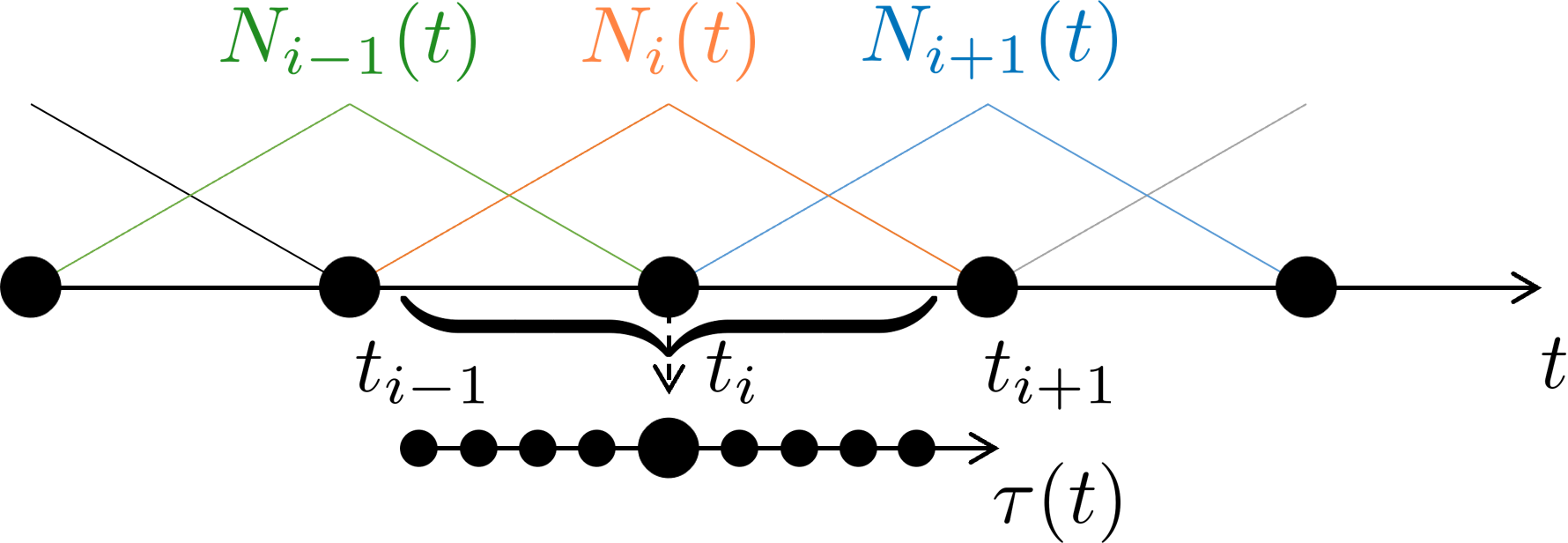}
\caption{Illustration of multi-scale discretization and used shape functions for the macro problem.}
\label{fig:MS_PU_idea}
\end{figure}

\begin{figure}[H] 
\centering
\includegraphics[width=0.5\textwidth]{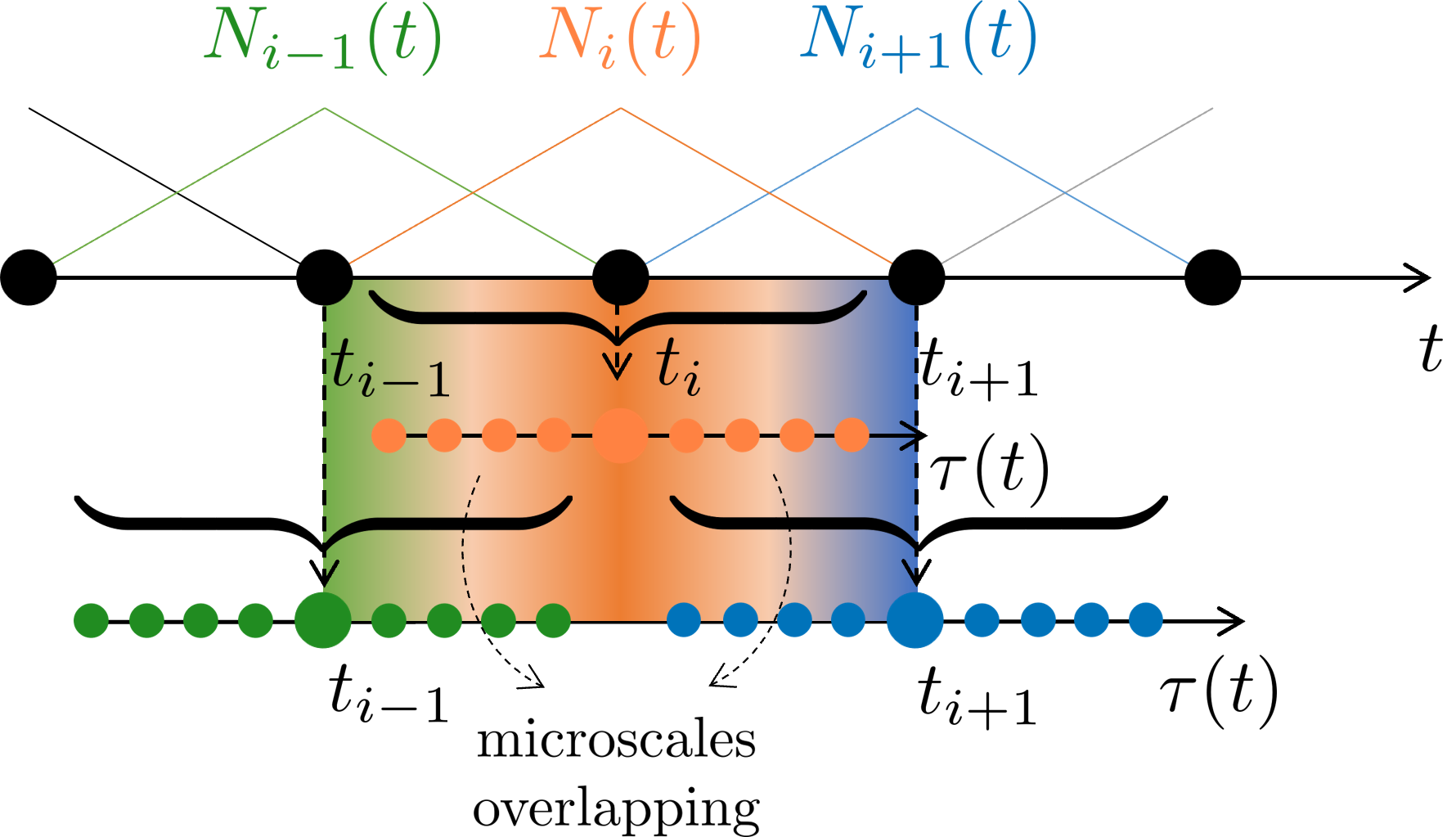}
\caption{Illustration of overlapping of microscales.}
\label{fig:MS_PU_overlapping}
\end{figure}

\begin{figure}[h!] 
\centering
\includegraphics[width=0.5\textwidth]{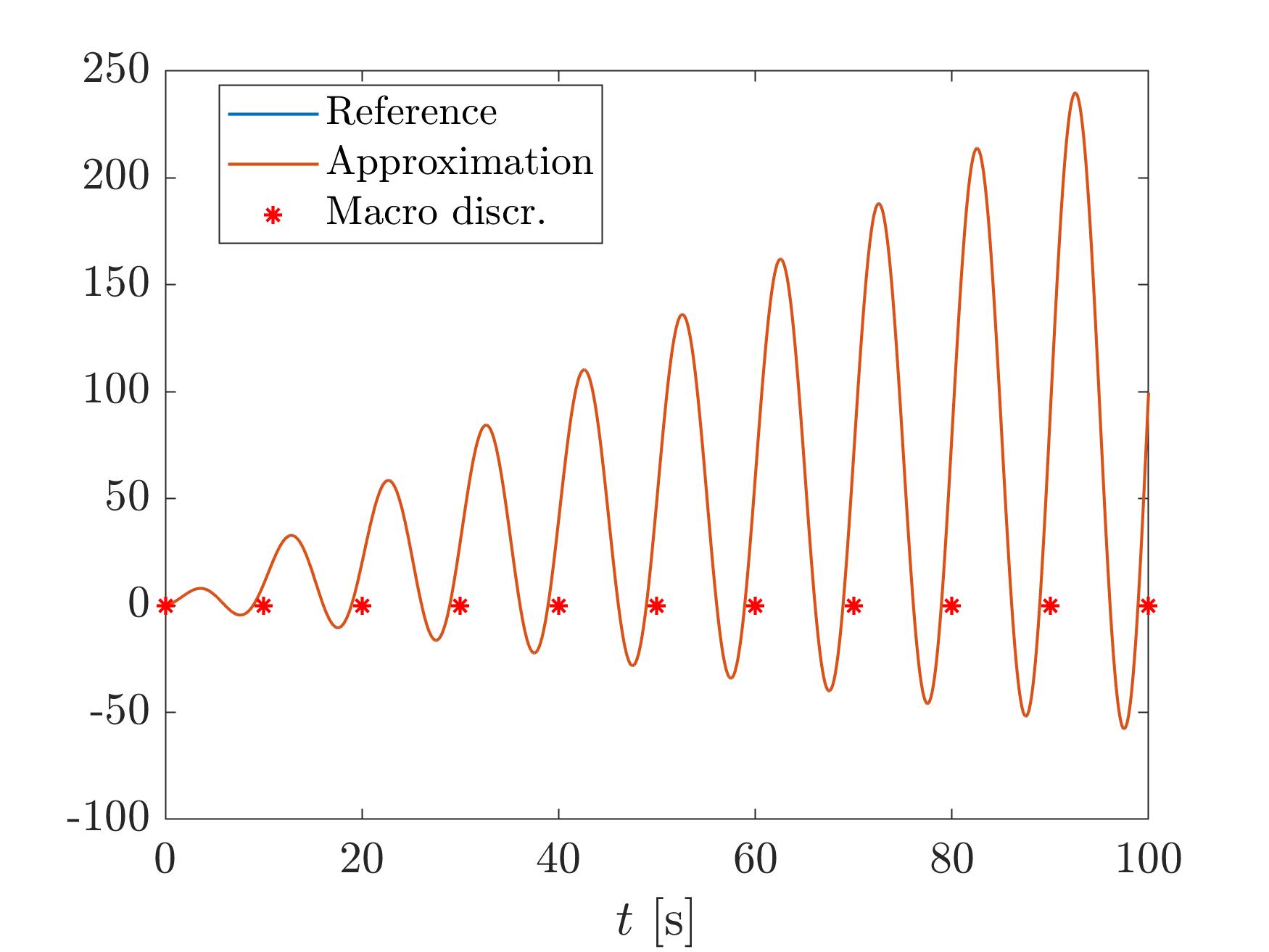}
\caption{Reference signal and multi-scale approximation.}
\label{fig:MS_comparison}
\end{figure}

\begin{figure}[h!] 
\centering
\begin{subfigure}{0.42\textwidth}
\includegraphics[width=\textwidth]{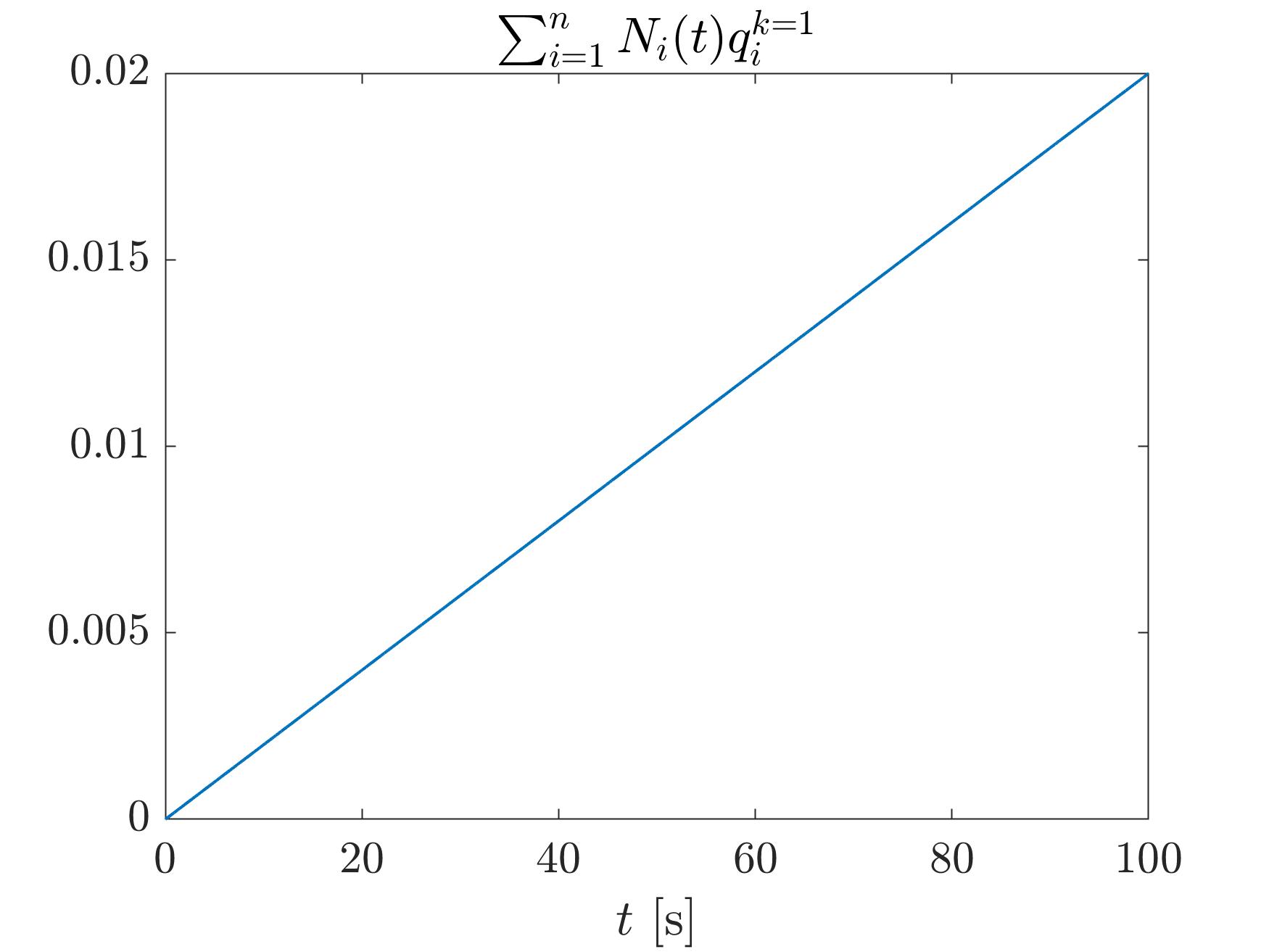}
\caption{Macro function.}
\label{fig:MS_macro}
\end{subfigure}
\begin{subfigure}{0.42\textwidth}
\includegraphics[width=\textwidth]{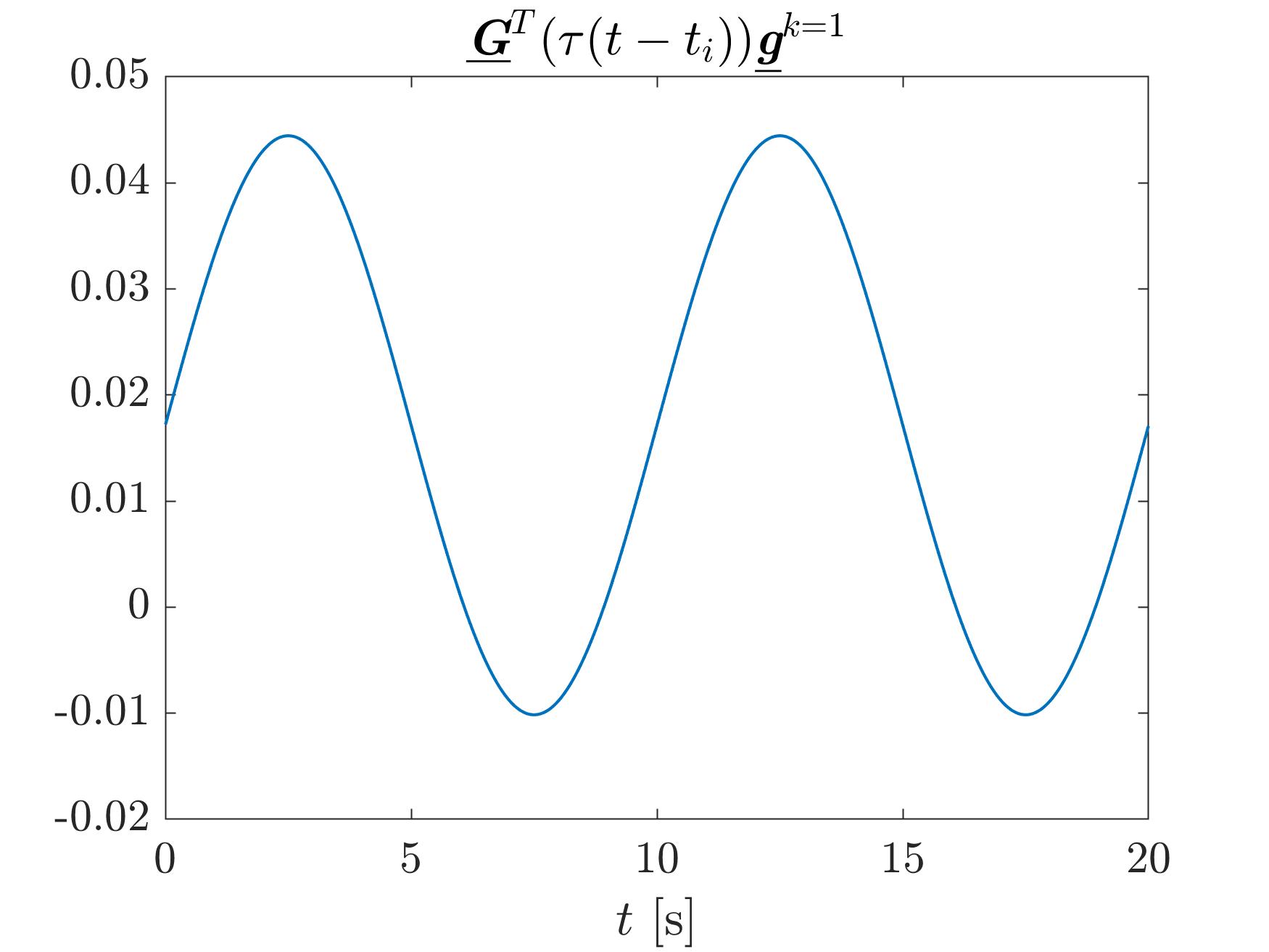}
\caption{Micro function.}
\label{fig:MS_micro}
\end{subfigure}
\caption{}
\end{figure}

From the approximation \eqref{eq:ms_approx} one can express the temporal derivative as follows:
\begin{equation}\label{eq:ms_approx_dt}
\dot{\pgdt}(t) \approx \sum_{k=1}^{\PGDm_{s}} \sum_{i=1}^{n} \left( \frac{\partial N_{i}(t)}{\partial t}q_{i}^{k} \SDvect{G}^{T}(\tau(t-t_{i})) \SDvect{g}^{k} + N_{i}(t)q_{i}^{k} \frac{\partial \SDvect{G}^{T}(\tau(t-t_{i}))}{\partial \tau} \frac{\partial \tau}{\partial t} \SDvect{g}^{k} \right)
\end{equation}
Figure \ref{fig:MS_PU_idea} shows the shape functions associated with both the macroscale (top) and the microscale (bottom). Notice how
a two-scale approach presents two meshes related to micro and macro scale, respectively.
Another important aspect of the multi-scale decomposition relying on the Partition of Unity is the interaction of the microscales betweeen contiguous macrointervals. Indeed, as shown in figure \ref{fig:MS_PU_overlapping}, the macrotime function $N_i$ is a hat function associated to the centroid $t_i$ of the microscale defined in the interval $[t_{i-1}, t_{i+1}]$. By construction, in the interval $[t_{i-1},t_i]$ there is an overlap of the effects of $N_{i-1}$ and $N_i$ ensuring the continuity of the approximation (since $N_i$ vanishes in $t_{i-1}$, while $N_{i-1}$ is maximum). The same occurs for $[t_i, t_{i+1}]$.
To illustrate the multi-scale approximation, figure \ref{fig:MS_comparison} represents a reference signal and its corresponding multi-scale approximation with an error of $0.04$ [\%] using the macro and micro functions presented in figures \ref{fig:MS_macro} and \ref{fig:MS_micro} respectively under the PU paradigm.
This temporal multi-scale approximation is used to boost the computation of the temporal functions of the PGD decomposition associated to the displacement $\lbrace \pgdt_{i}(t) \rbrace_{i=1}^{\PGDm}$ and internal variables $\lbrace \phi^{[j]}_{i}(t) \rbrace_{i=1}^{\PGDmz^{[j]}}$ of \eqref{eq:PGD_displacement} and \eqref{eq:PGD_int_var} respectively, by introducing its approximation into the problems \eqref{eq:eq_solv_disp} and \eqref{eq:int_var_min} and solving their corresponding macro and micro functions.

\subsubsection{Dealing with transient behavior in the time function to be approximated}\label{sec:lamda_c}

It should be remembered here that a multi-scale approximation works very well when the signal to be approximated can be correctly separated into two dynamics, the macro consisting of the slow one and the micro the fast one. In this sense, when the function to be approximated has transient behaviors, the multi-scale decomposition will lose its efficiency and many modes will have to be computed to simulate the transient.

In this sense, this paper proposes to apply the multi-scale approximation in conjunction with a classical or single-scale approximation. On the one hand, the classical approximation is applied to approximate the transient behavior of the signal (present in the first instants of time), and on the other hand the multi-scale approximation is applied to reproduce correctly the forced regime. 

Therefore, the final approximation can be written as follows:
\begin{equation}\label{eq:classic_plus_MS}
\pgdt(t) \approx  \Pi_{0,T_{c}}(t) \pgdt_{\text{c}}(t) + \Pi_{T_{c},T}(t) \sum_{k=1}^{\PGDm_{s}} \sum_{i=1}^{n} N_{i}(t)q_{i}^{k} \SDvect{G}^{T}(\tau(t-t_{i})) \SDvect{g}^{k}
\end{equation}
with $\Pi_{a,b}(t)$ the boxcar function defined as follows:
\begin{equation}
\Pi_{a,b}(t) = H(t-a) - H(t-b)
\end{equation}
with $H(t-a)$ the Heaviside step function defined as follows:
\begin{equation}
H(t-a) = \begin{cases}
1 \qquad \text{if} \ t \geq  a \\
0 \qquad \text{if} \ t < a
\end{cases}
\end{equation}
Here $\pgdt_{c}(t)$ the solution computed using classical FEM approach in time but only defined between $(0,T_{c})$. The time $T_{c}$ is a hyper-parameter of the method, empirical results show that a time $T_{c}$ equal to $2$ macro elements is sufficient to eliminate the transient component of the solution to be approximated.

\section{Numerical examples}\label{sec:numerical_example}

Here we consider a numerical example which consists on a 1D polymer bar. The dimensions of the bar correspond to: length $L=5\times10^{-3} [m]$ , and a square sectional area $A = 2.5\times10^{-7} [m^2]$. The considered polymer is polypropylene and its properties correspond to $E_v = 1.2 \ [\text{GPa}]$ and $E_{\infty}^{r} = 1 \ [\text{GPa}]$.

\begin{figure}[H] 
\centering
\begin{subfigure}{0.45\textwidth}
\includegraphics[width=\textwidth]{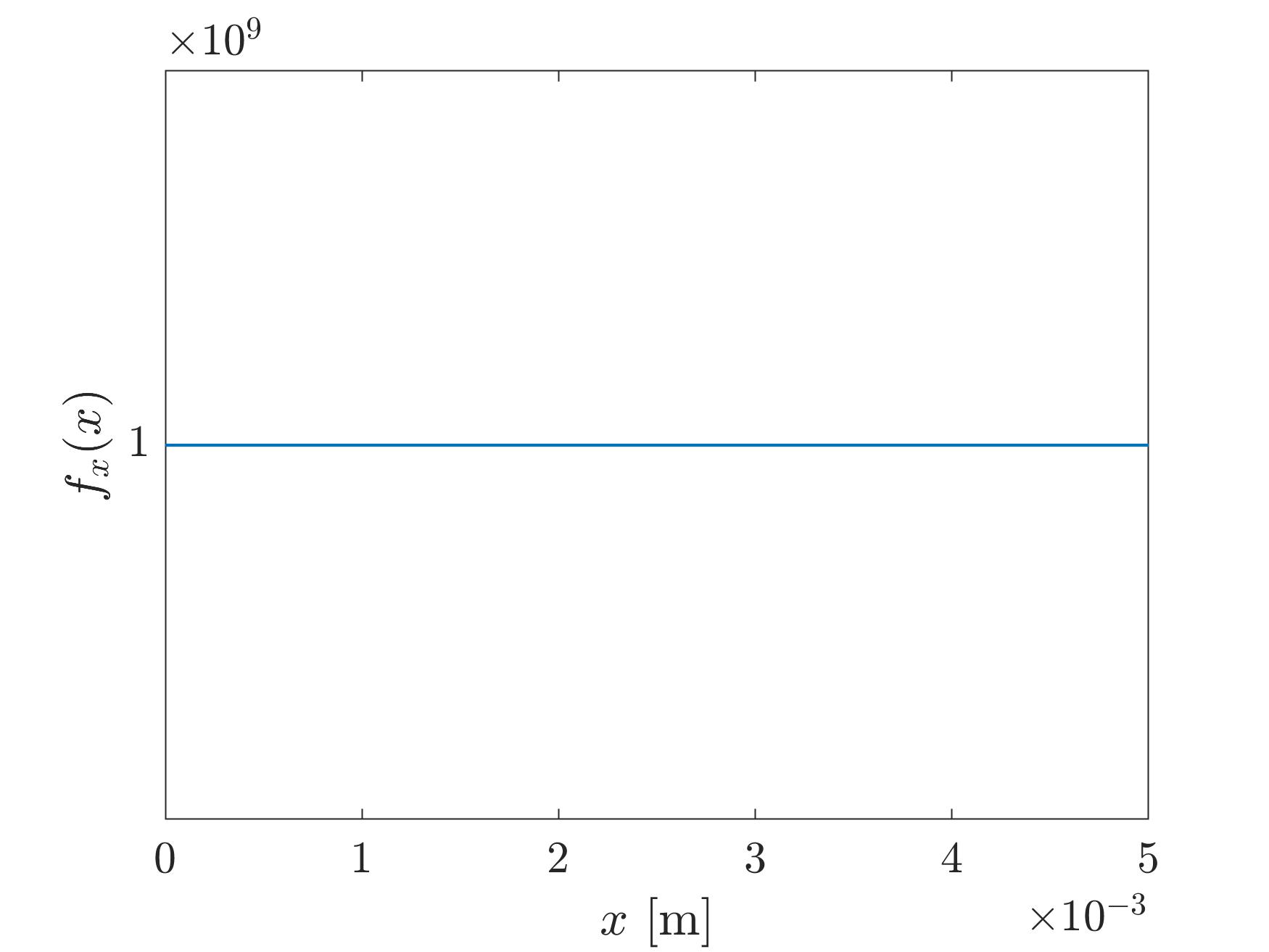}
\caption{Spatial function $f_{x}(x)$.}\label{fig:space_force}
\end{subfigure}
\begin{subfigure}{0.45\textwidth}
\includegraphics[width=\textwidth]{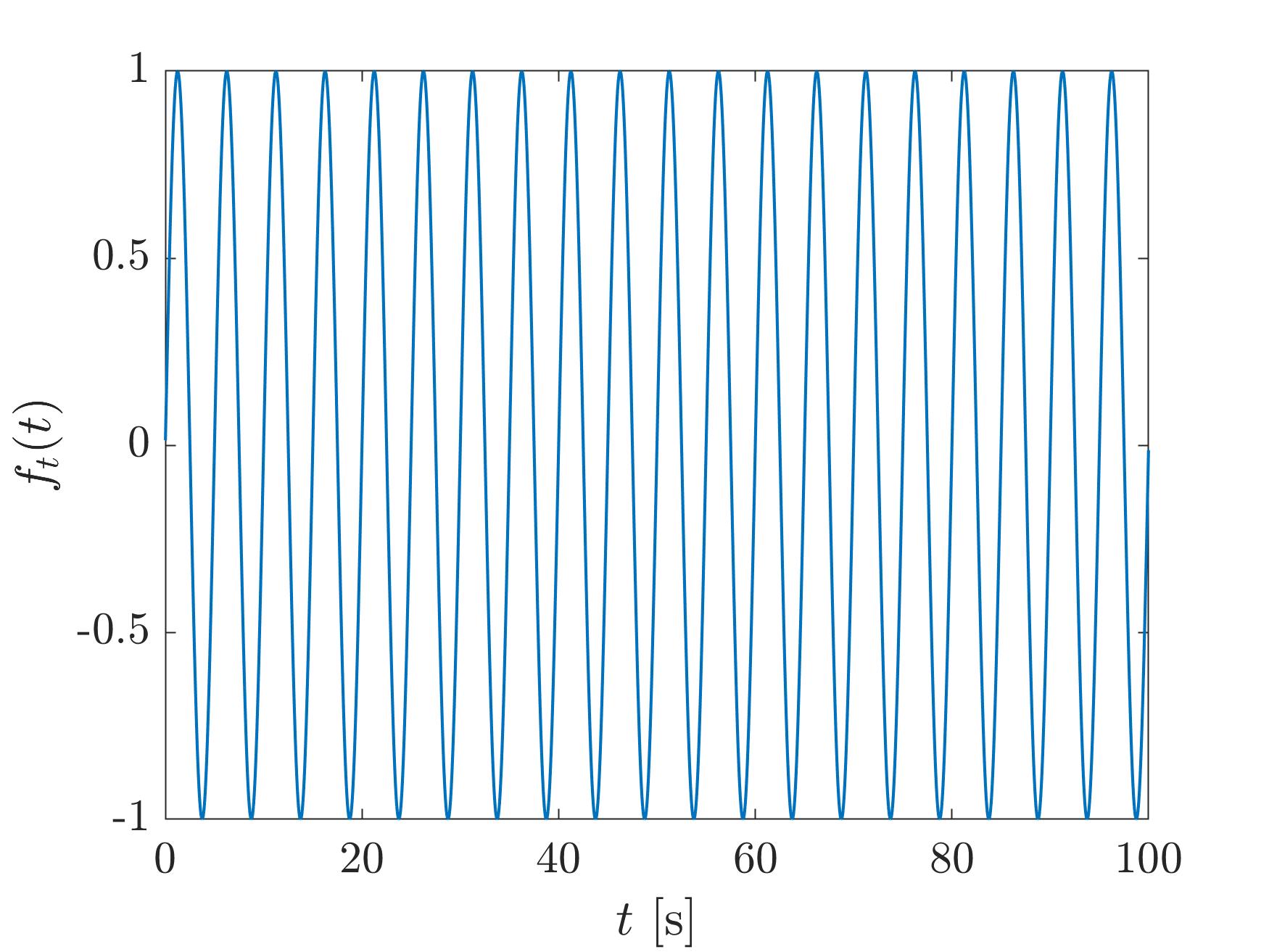}
\caption{Temporal function $f_{t}(t)$.}\label{fig:temporal_force}
\end{subfigure}
\caption{}
\end{figure}

The load considered here consists on a volumetric one, which is imposed along the whole bar. This force is represented as follows:
\begin{equation}
f(x,t) = f_{x}(x) f_{t}(t) \quad [N/m]
\end{equation}
where $f_{x}(x)$ and $f_{t}(t)$ are illustrated in figures \ref{fig:space_force} and \ref{fig:temporal_force} respectively.
In what follows, a stop criteria of $2$ [\%] is imposed for the low-rank approximation of displacement and internal variables for the PGD resolution as well as the PGD multi-scale in time following the expressions presented in sections \ref{sec:error_crit_PGD_u} and \ref{sec:error_crit_PGD_iv}.

The time interval considered corresponds to 100 [s], for this problem. Linear shape functions are considered for the time resolution both for the single-scale and for the macro and micro functions of the multi-scale resolution. The single-scale discretization used considers $2001$ nodal values in time. On the other hand, the use of the multi-scale approximation allows to drastically reduce these DOFs, in fact, this approximation considers two groups, the degrees of freedom associated to the solution of the macro problem and the micro one. The macro problem considers $21$ DOFs while the micro $201$ DOFs, which is a reduction of $89$ [\%] of degrees of freedom for the solution of the temporal problem at each iteration of the solver.

\subsection{One internal variable}

Here only $1$ internal variable is considered, in order to better understand and analyse the numerical results. For this example a relaxation time of $\tau = 5 [s]$ is considered as well as a weight $p_{1} = 0.025$.

\begin{figure}[H] 
\centering
\begin{subfigure}[t]{0.45\textwidth}
\includegraphics[width=\textwidth]{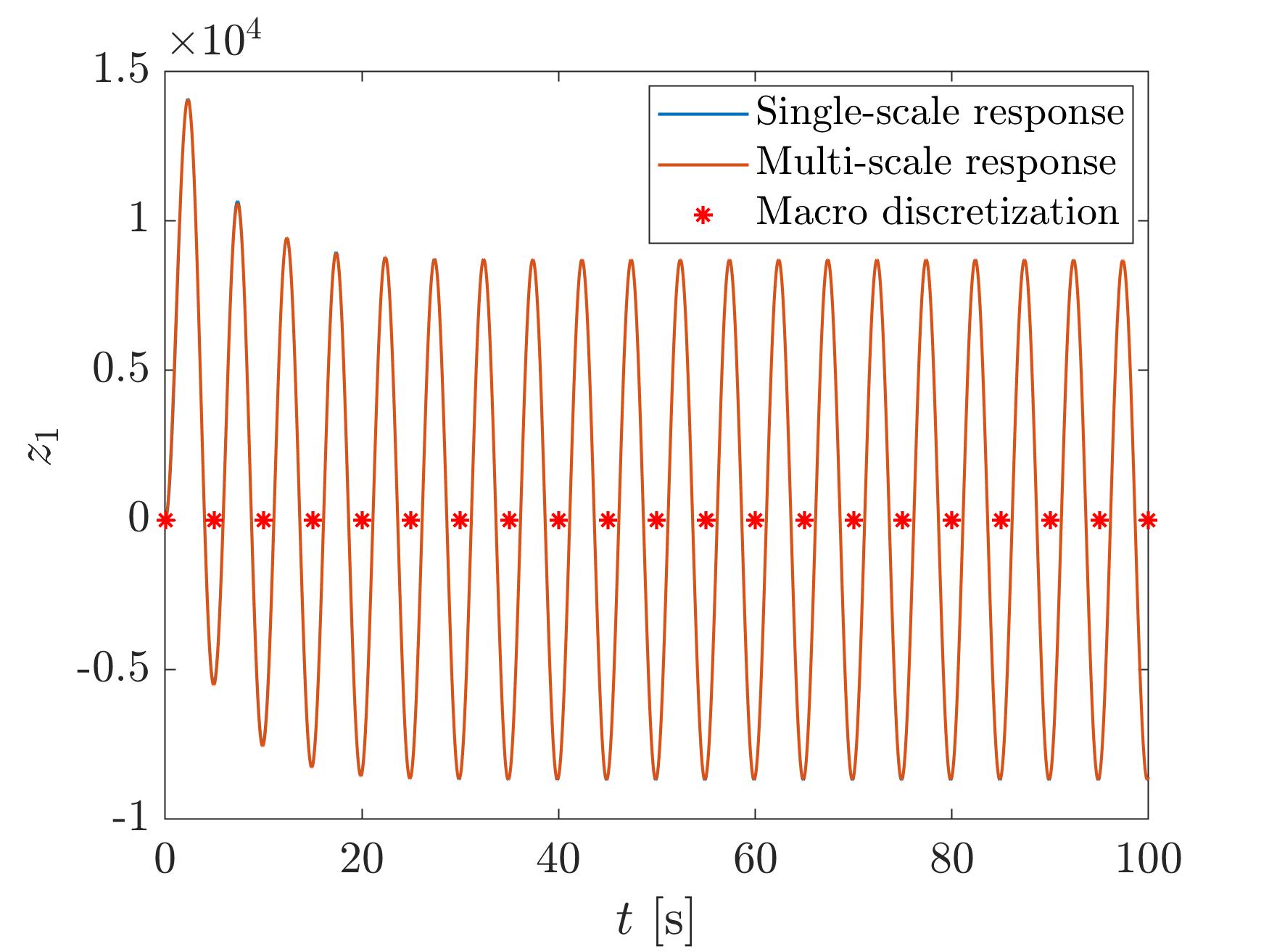}
\caption{Reference temporal mode and its multi-scale approximation.}
\label{fig:first_mode_intvar}
\end{subfigure}
\begin{subfigure}[t]{0.45\textwidth}
\includegraphics[width=\textwidth]{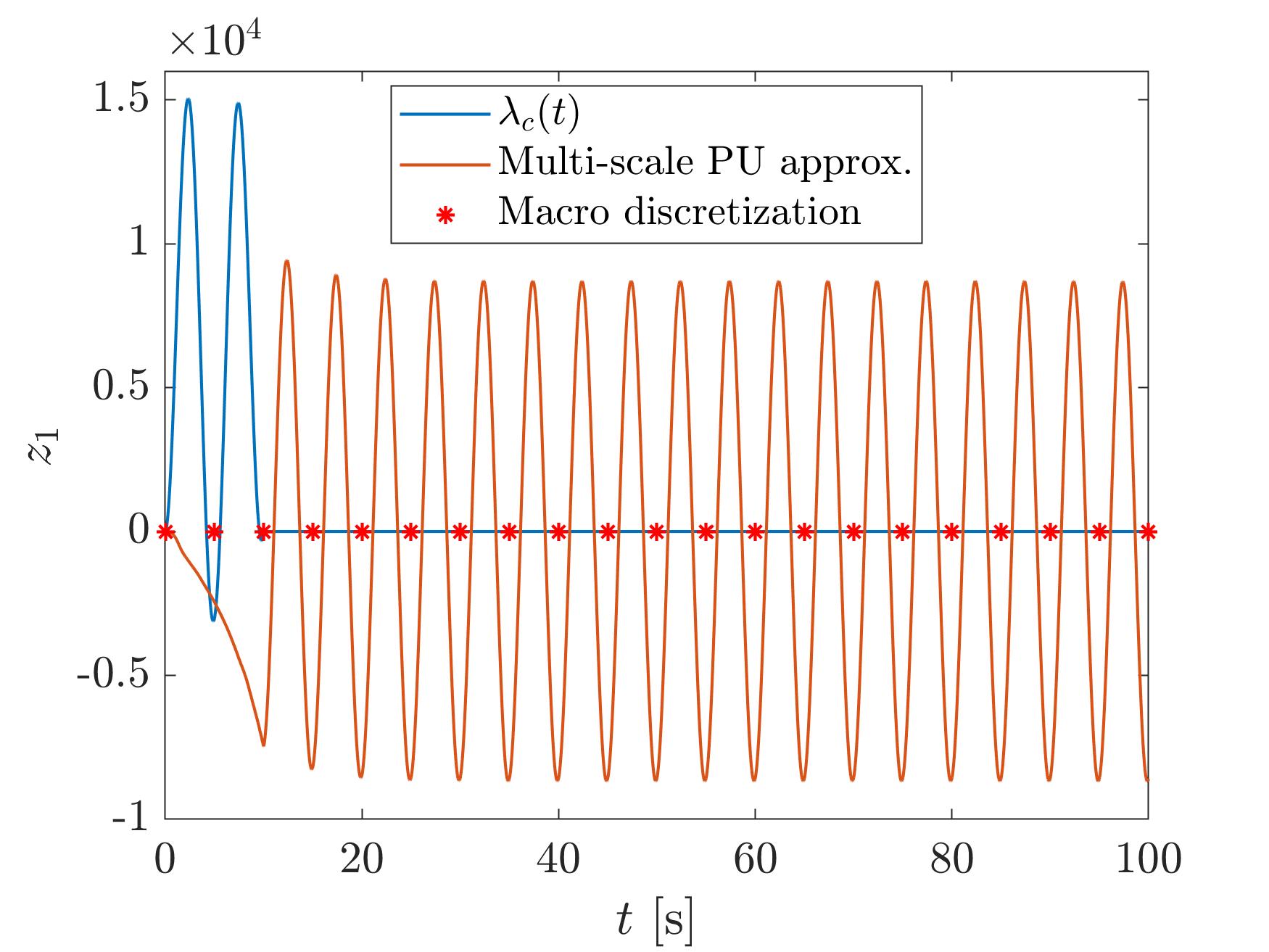}
\caption{Contribution of the temporal function $\pgdt_{c}(t)$ with respect to the multiscale approximation.}
\label{fig:trans_func}
\end{subfigure}
\caption{}
\end{figure}

\begin{figure}[H] 
\centering
\begin{subfigure}{0.45\textwidth}
\includegraphics[width=\textwidth]{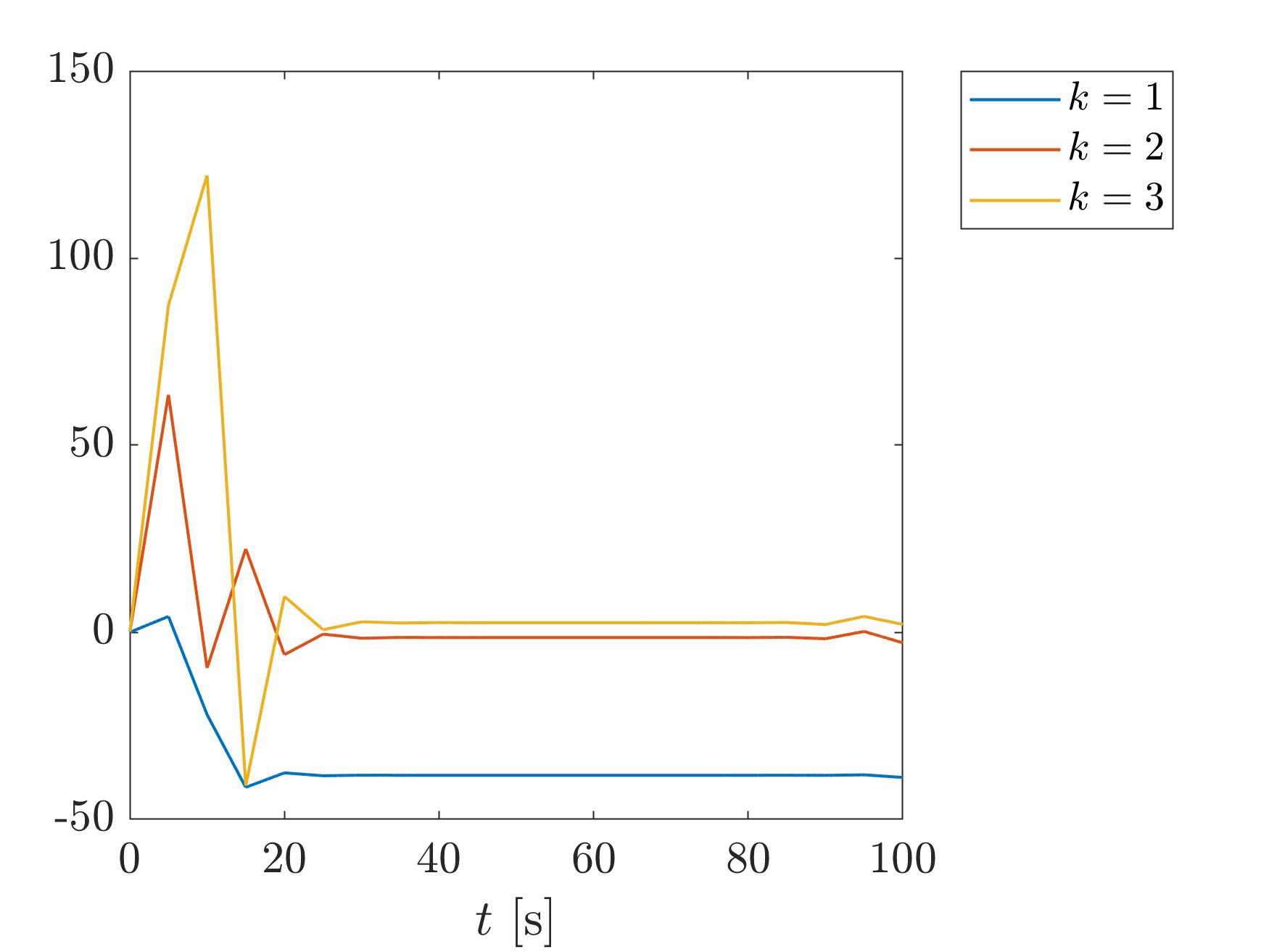}
\caption{Macro functions.}\label{fig:macro_modes}
\end{subfigure}
\begin{subfigure}{0.45\textwidth}
\includegraphics[width=\textwidth]{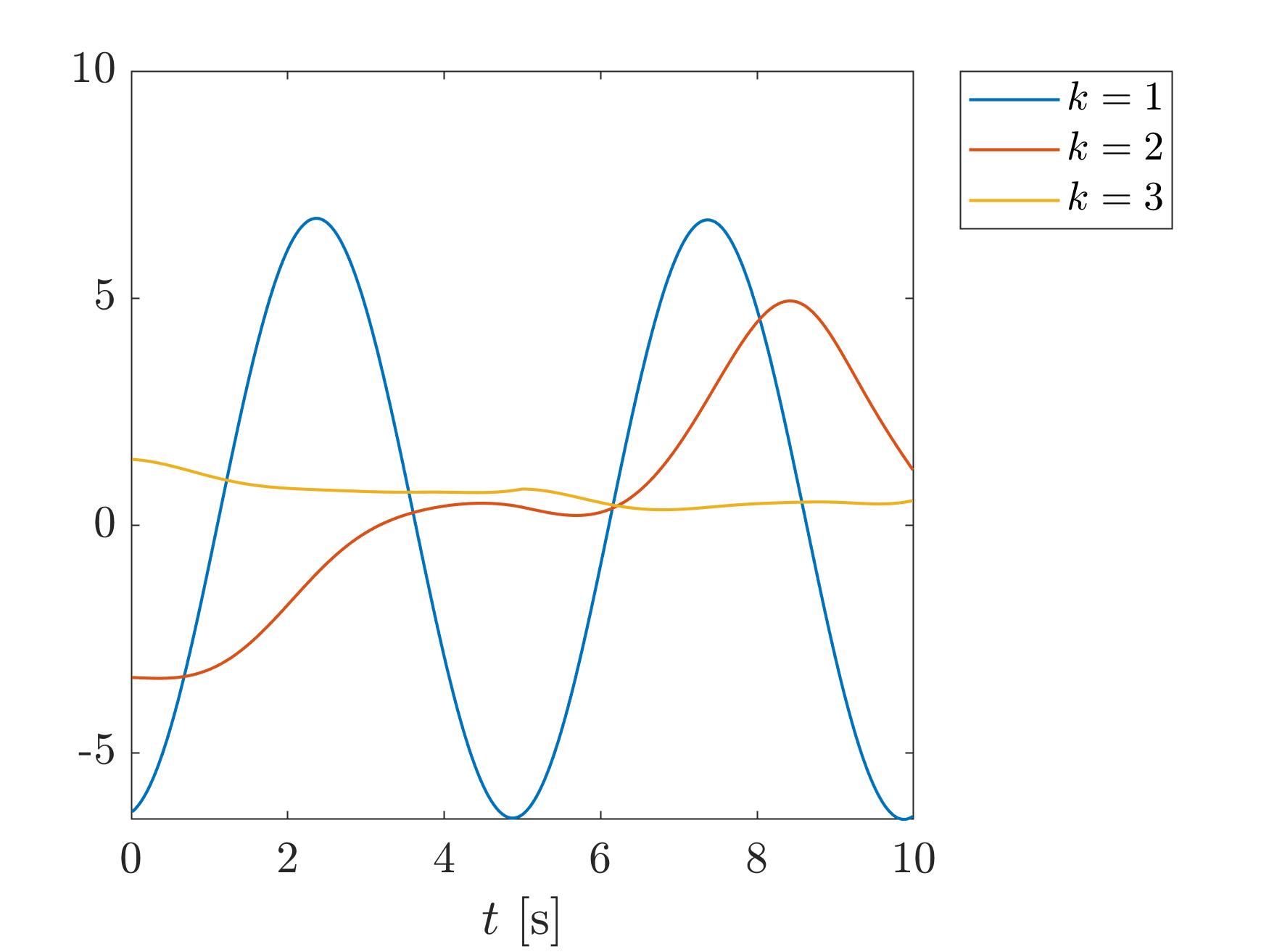}
\caption{Micro functions.}\label{fig:micro_modes}
\end{subfigure}
\caption{}
\end{figure}

\begin{figure}[H] 
\centering
\begin{subfigure}{0.45\textwidth}
\includegraphics[width=\textwidth]{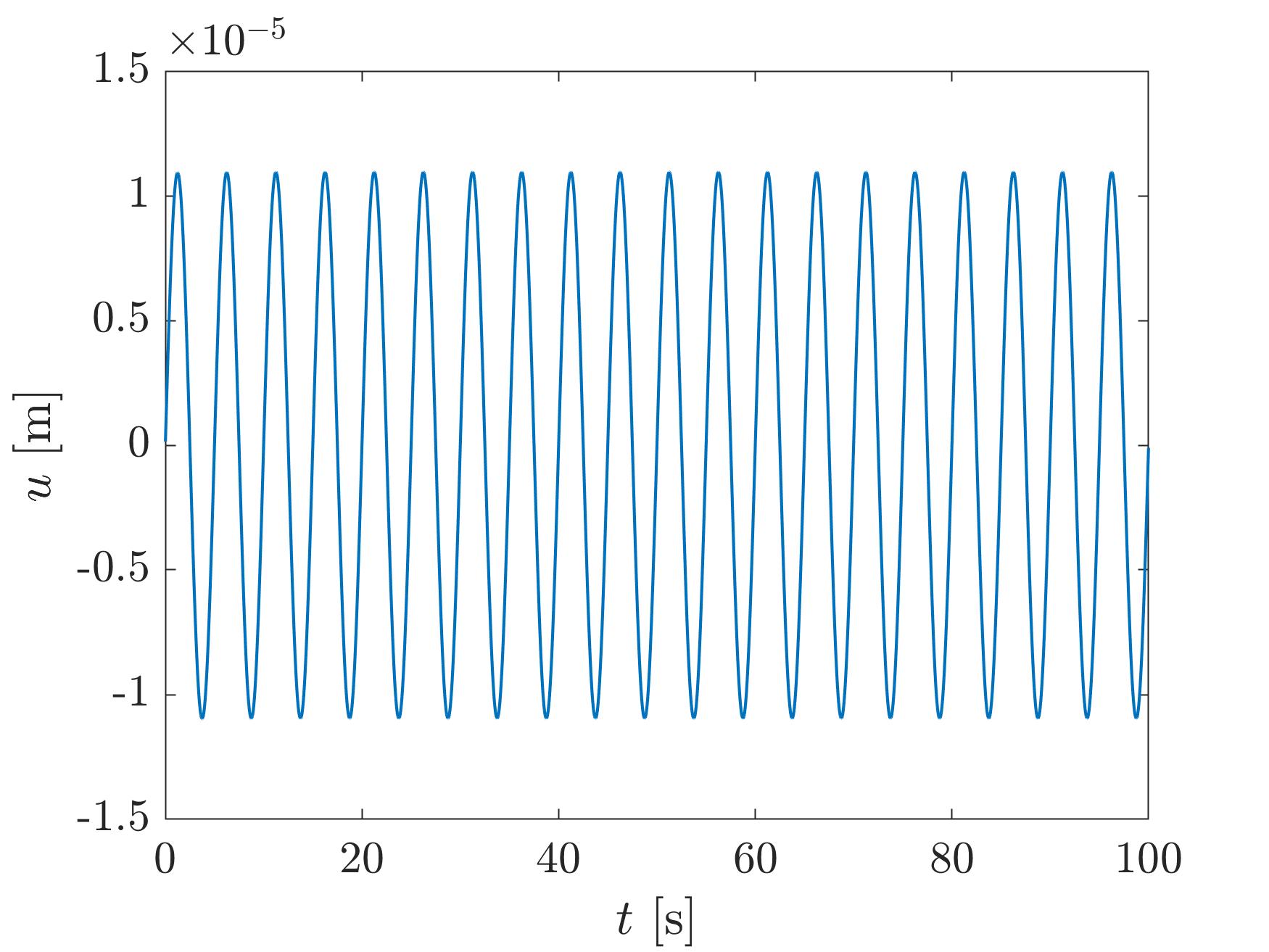}
\caption{Single-scale solution.}
\end{subfigure}
\begin{subfigure}{0.45\textwidth}
\includegraphics[width=\textwidth]{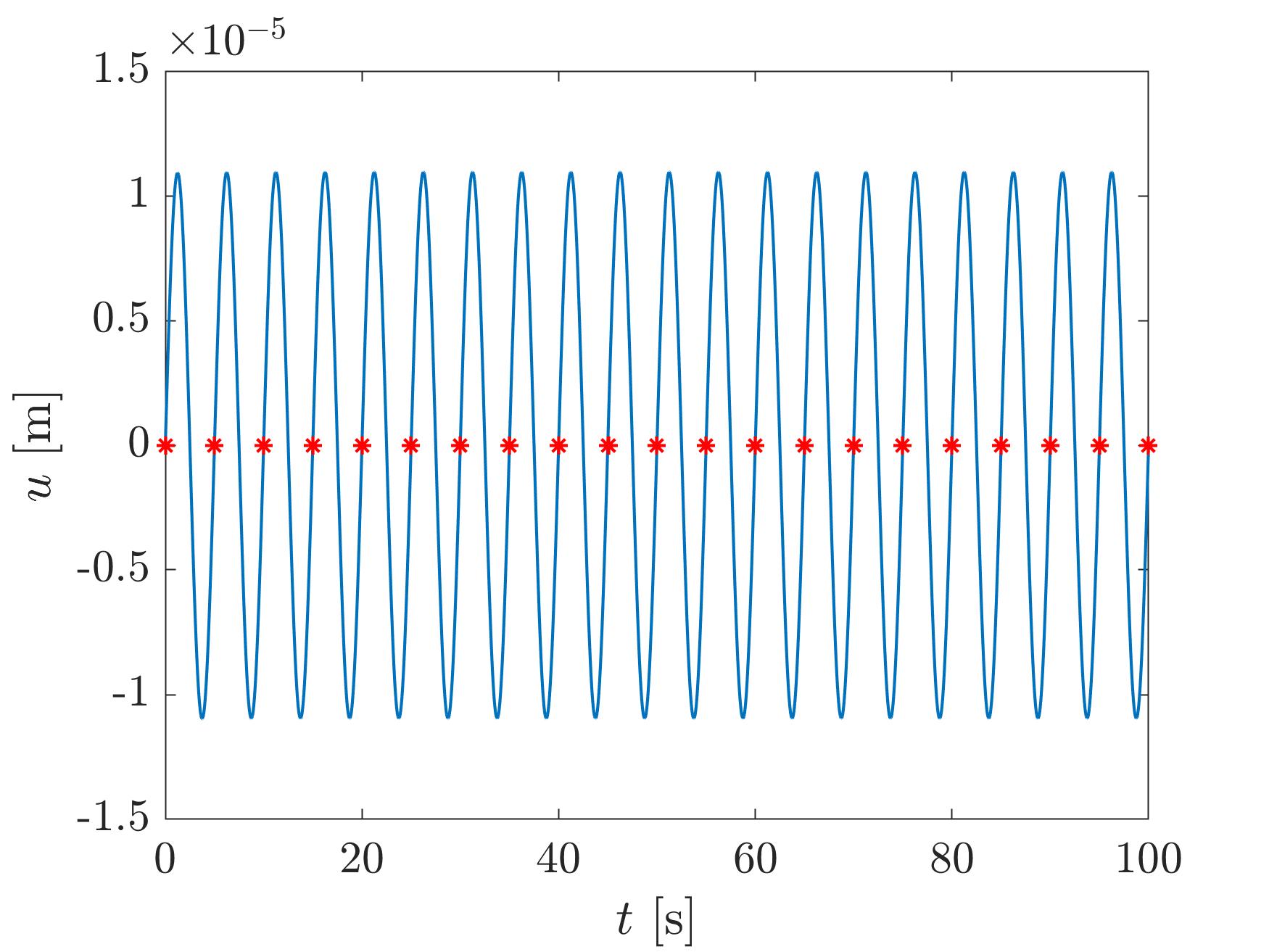}
\caption{Multi-scale solution.}
\end{subfigure}
\caption{Displacement at the middle of the bar of the converged solution (red dots correspond to the macro discretization).}\label{fig:disp_comp}
\end{figure}

\begin{figure}[H] 
\centering
\begin{subfigure}{0.45\textwidth}
\includegraphics[width=\textwidth]{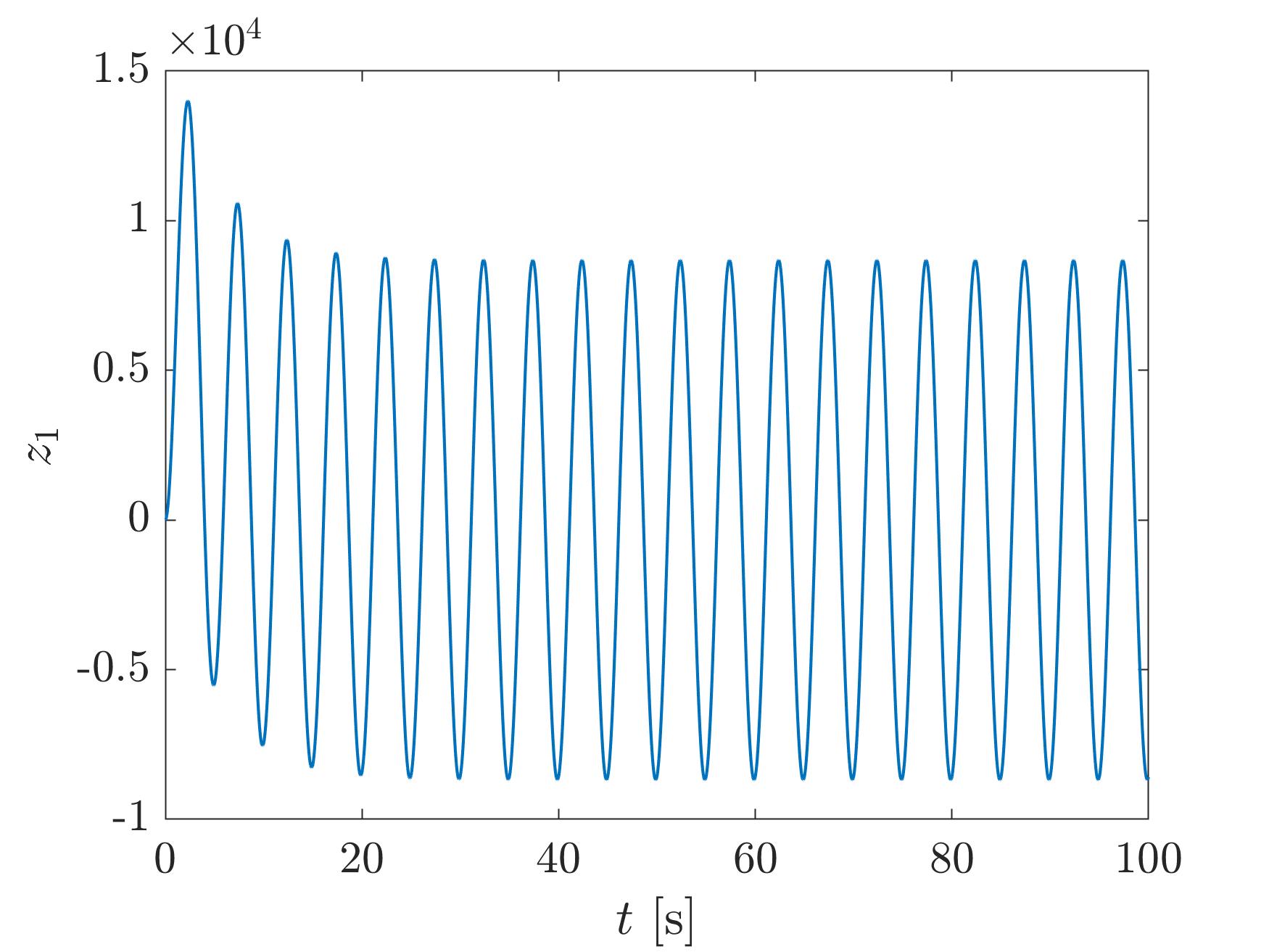}
\caption{Single-scale solution.}
\end{subfigure}
\begin{subfigure}{0.45\textwidth}
\includegraphics[width=\textwidth]{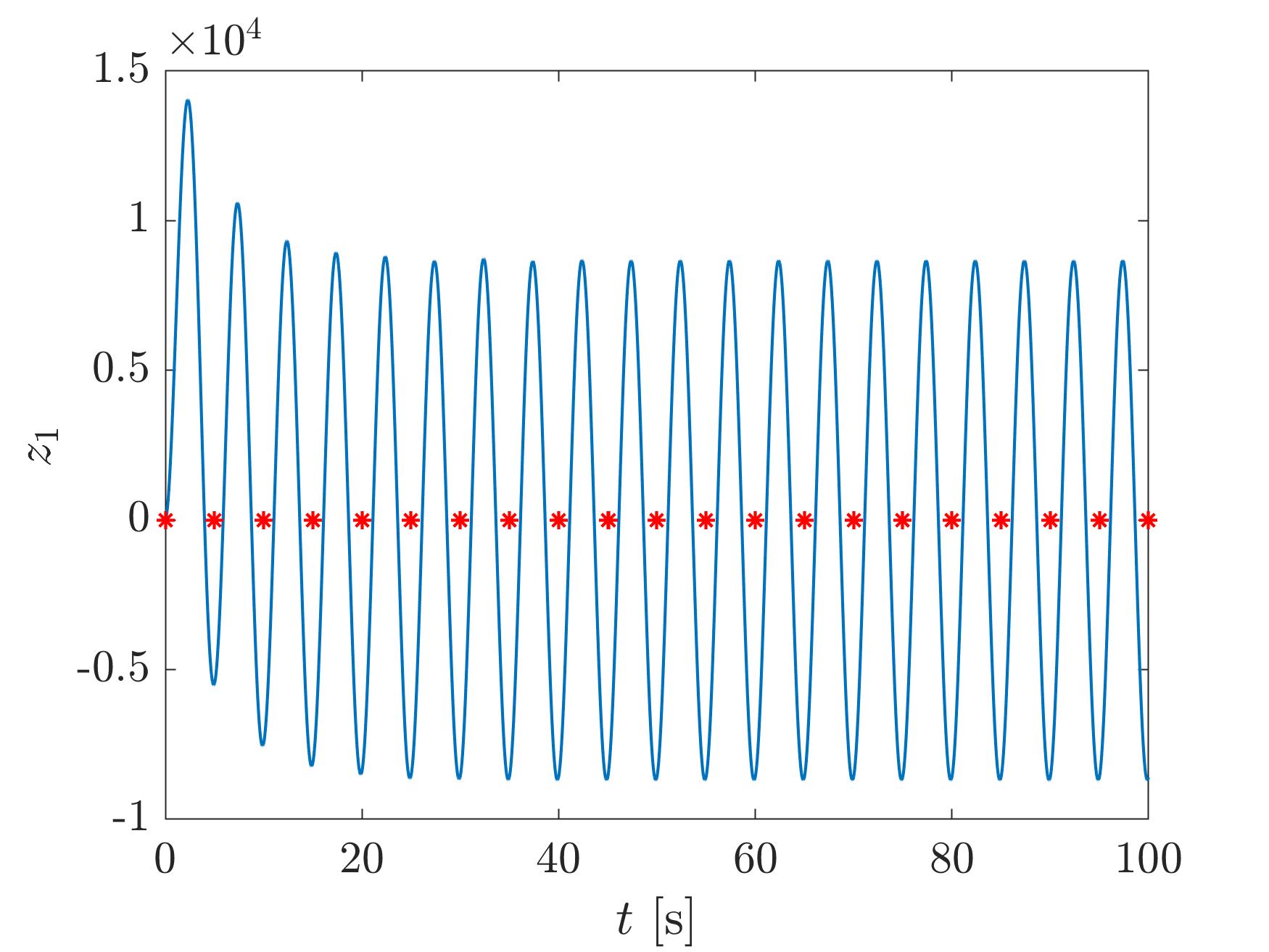}
\caption{Multi-scale solution.}
\end{subfigure}
\caption{Internal variable at the middle of the bar of the converged solution (red dosts corresponds to the macro discretization).}\label{fig:int_var_comp}
\end{figure}

\begin{figure}[H] 
\centering
\begin{subfigure}{0.45\textwidth}
\includegraphics[width=\textwidth]{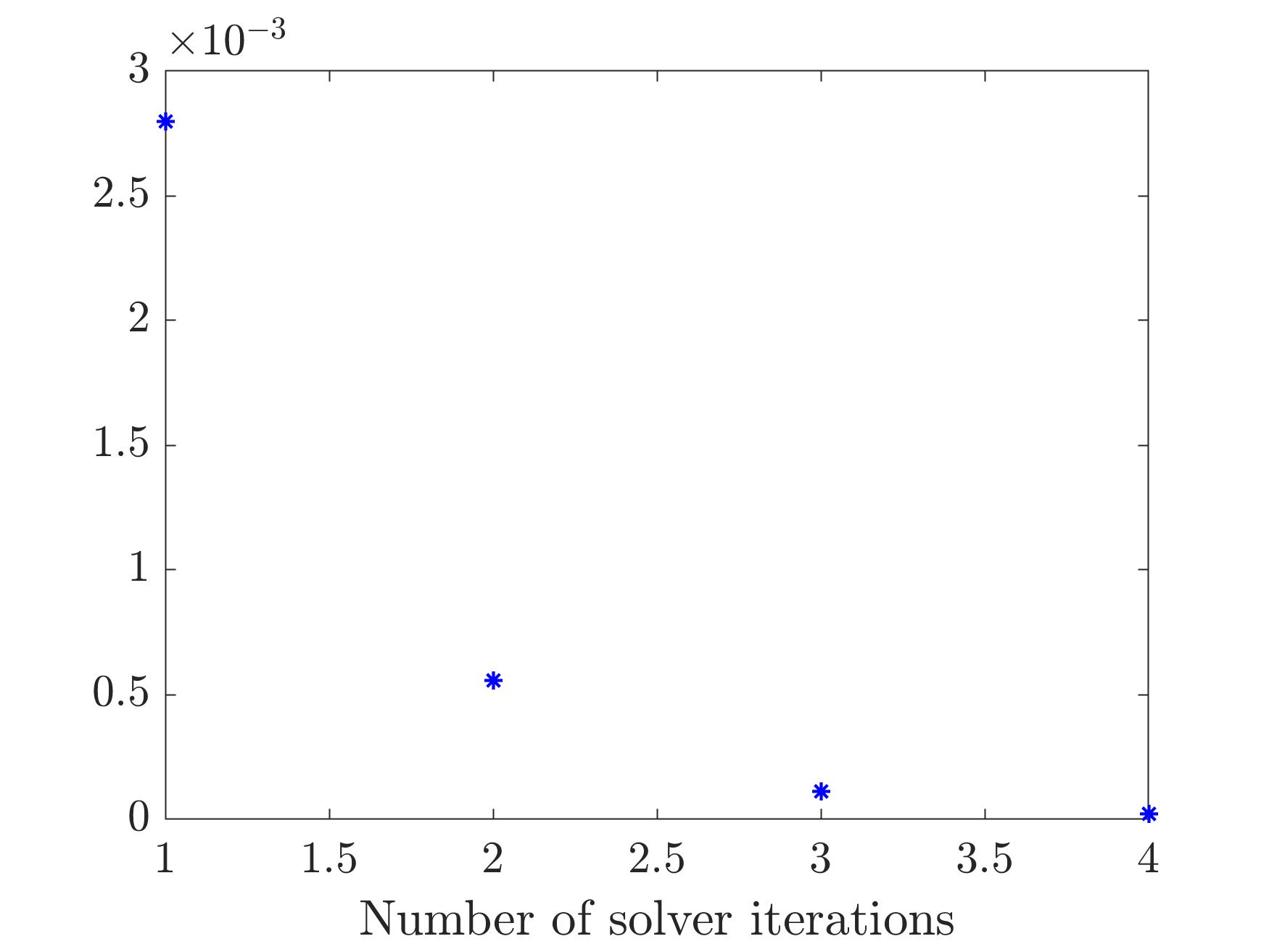}
\caption{Single-scale solution stagnation indicator.}
\end{subfigure}
\begin{subfigure}{0.45\textwidth}
\includegraphics[width=\textwidth]{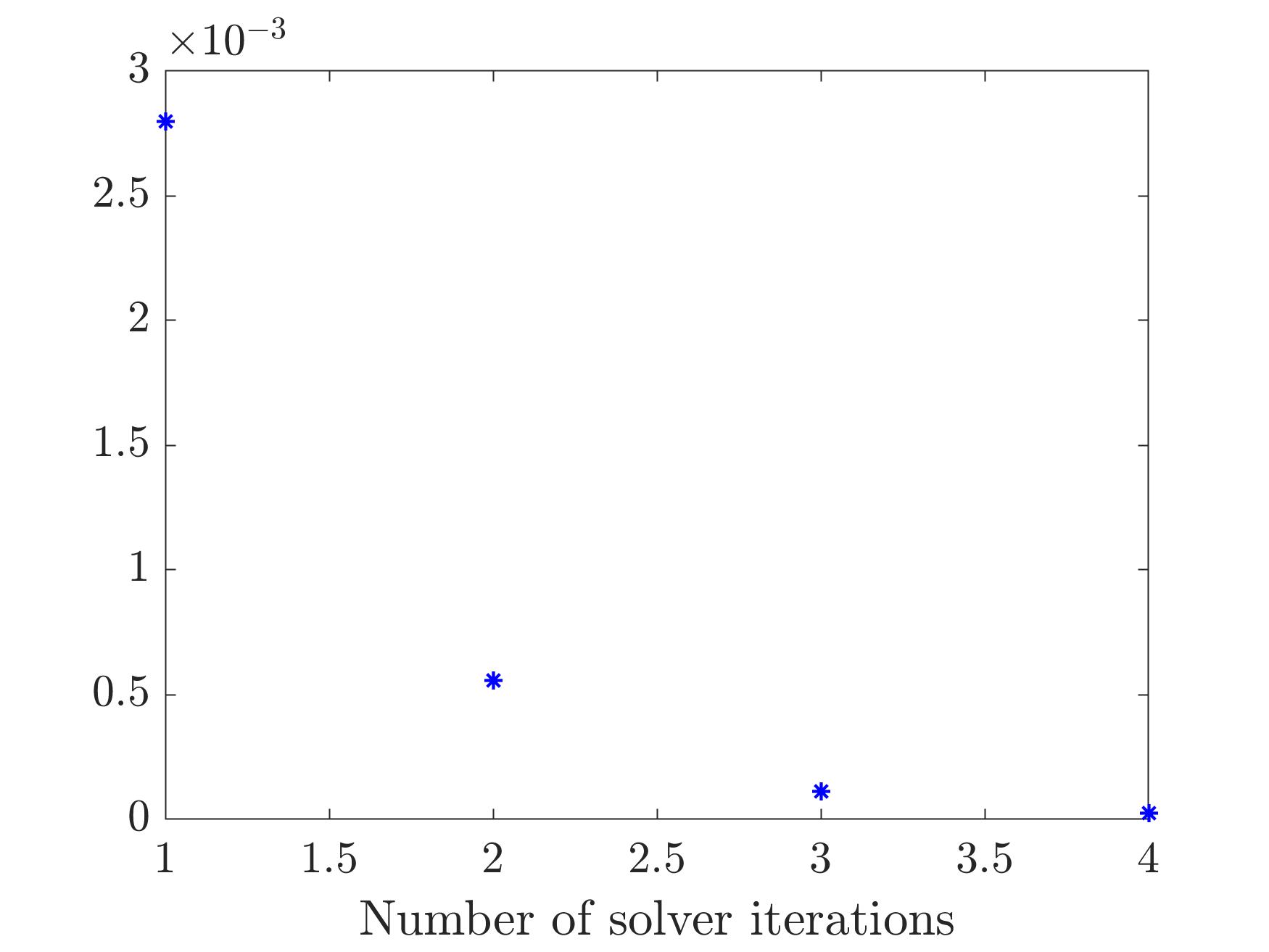}
\caption{Multi-scale solution stagnation indicator.}
\end{subfigure}
\caption{Stagnation of solvers.}\label{fig:stagnation}
\end{figure}

Figure \ref{fig:first_mode_intvar} presents the multi-scale approximation and the reference for the determination of the first temporal mode of the internal variable at the first solver iteration, where an approximation error of $0.52$ [\%] is obtained. In addition, figure \ref{fig:trans_func} illustrates the  solution $\pgdt_{c}(t)$ computed to describe the transient behavior of the first mode of the temporal PGD function related to the internal variable (see section \ref{sec:lamda_c}).
%
%
%
%
While figures \ref{fig:macro_modes} and \ref{fig:micro_modes} show the first 3 macro and micro functions respectively.
Figures \ref{fig:disp_comp} and \ref{fig:int_var_comp} present a comparison of the results for the displacement and internal variable respectively at convergence for both single and multi-scale approach.
The use of the multi-scale approximation induces in some cases an extra number of iterations needed to be carried out in order to converge. Figure \ref{fig:stagnation} presents the rate of convergence of the solver using both single and multi-scale resolution in time.

\subsection{50 internal variables}

\begin{figure}[h!] 
\centering
\includegraphics[width=0.5\textwidth]{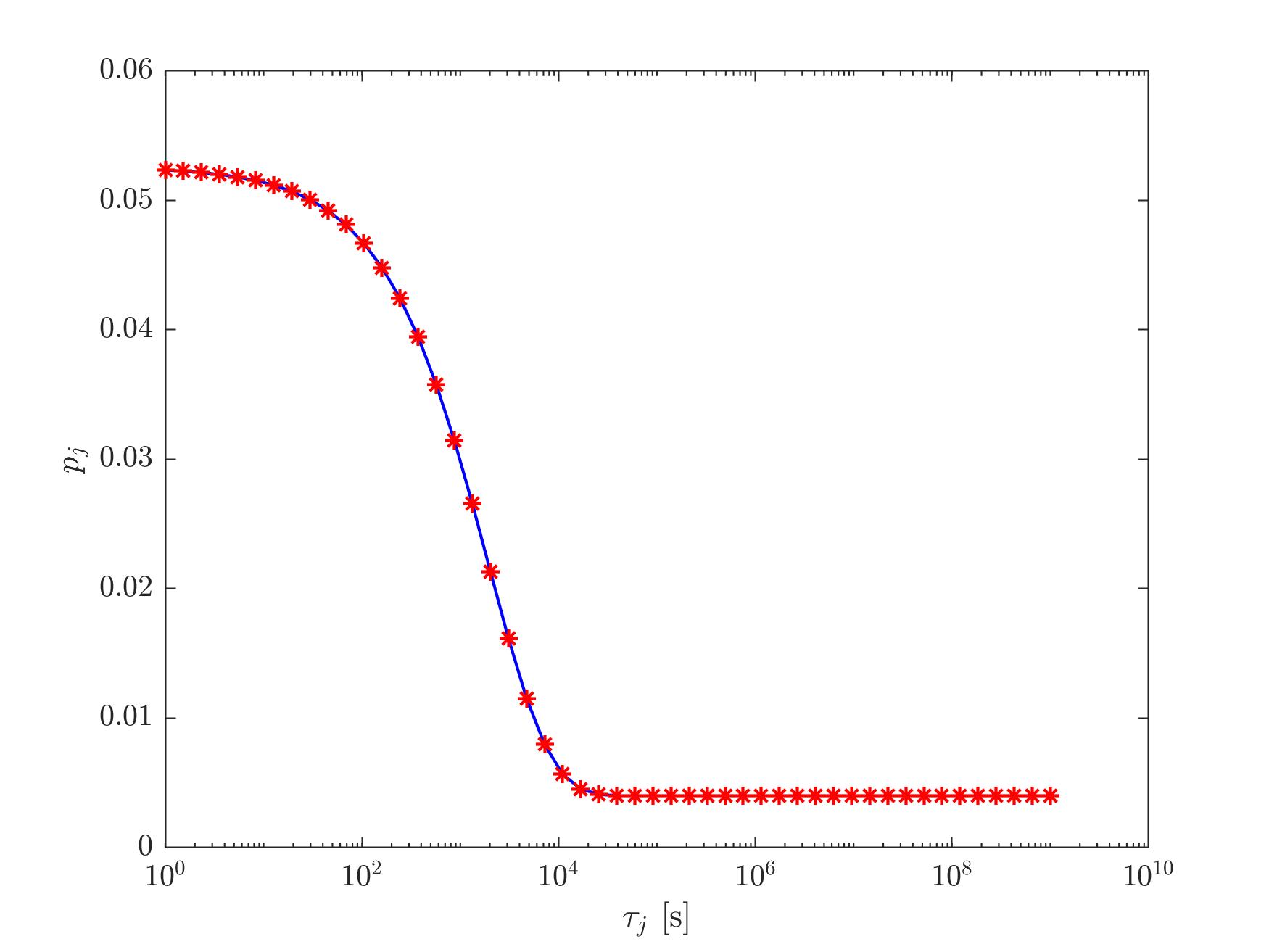}
\caption{Spectrum of weight distribution versus relaxation times.}
\label{fig:weights_distr}
\end{figure}

Here, $50$ internal variables are considered each of them with different relaxation times. The spectrum of the weight distribution with respect to the relaxation times considered are illustrated in figure \ref{fig:weights_distr}.

\begin{figure}[H] 
\centering
\begin{subfigure}{0.45\textwidth}
\includegraphics[width=\textwidth]{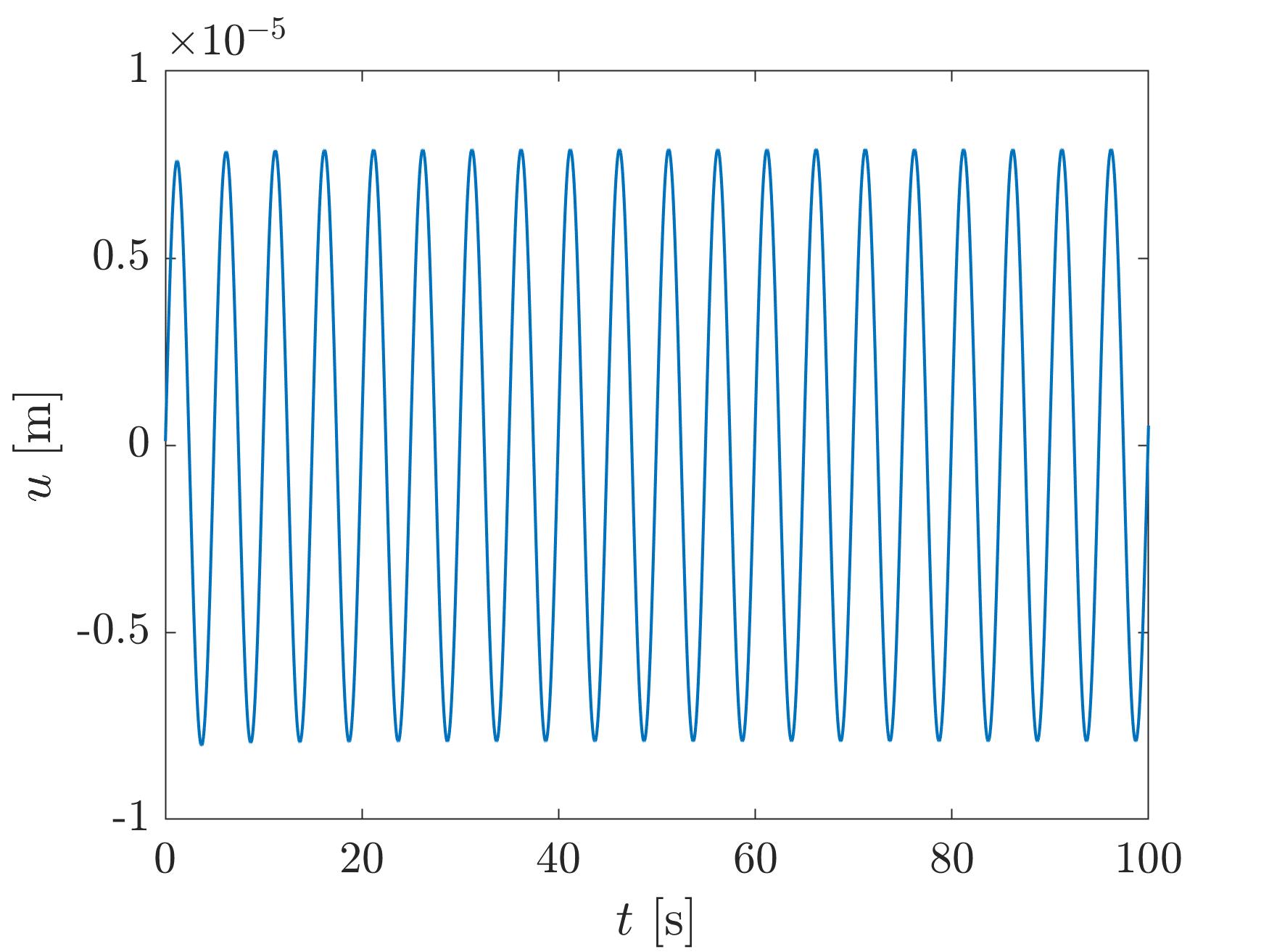}
\caption{Single-scale solution.}\label{fig:disp_comp_50}
\end{subfigure}
\begin{subfigure}{0.45\textwidth}
\includegraphics[width=\textwidth]{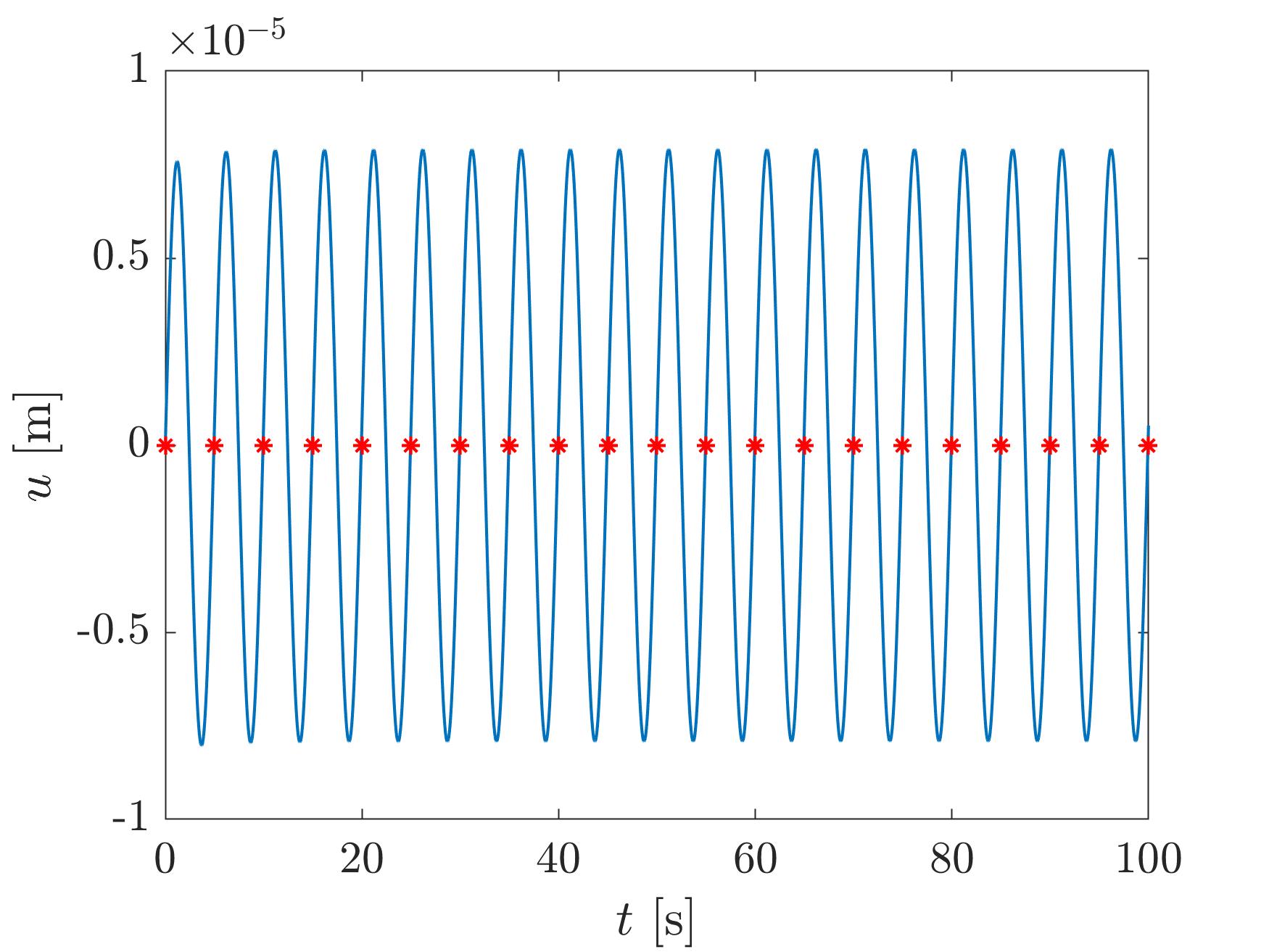}
\caption{Multi-scale solution.}
\label{fig:disp_comp_ms_50}
\end{subfigure}
\caption{Displacement at the middle of the bar of the converged solution (red dosts corresponds to the macro discretization).}
\end{figure}

\begin{figure}[H] 
\centering
\begin{subfigure}{0.45\textwidth}
\includegraphics[width=\textwidth]{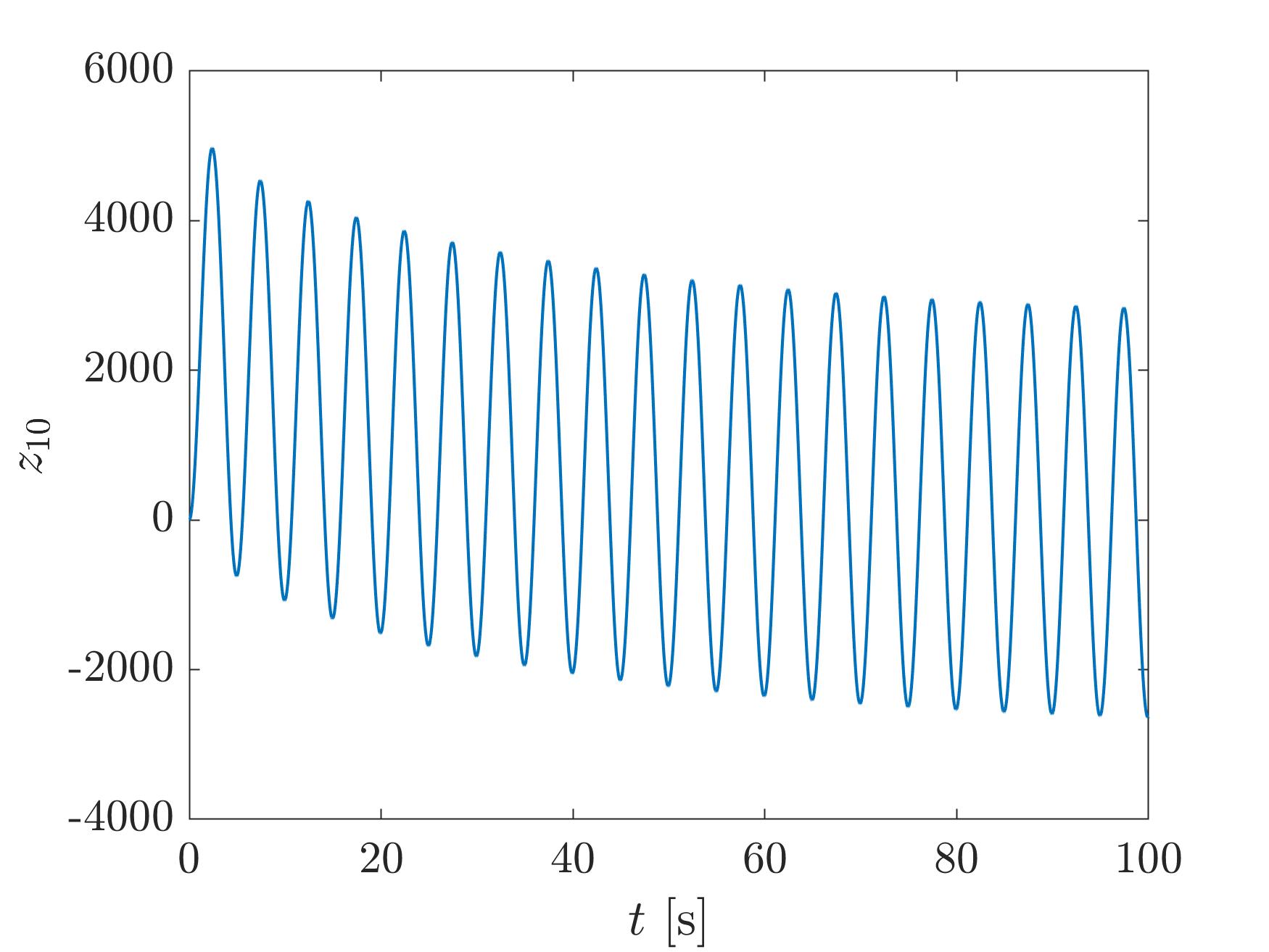}
\caption{$10^{\text{th}}$ internal variable solution using a single-scale approach.}\label{fig:iv_10_classic}
\end{subfigure}
\begin{subfigure}{0.45\textwidth}
\includegraphics[width=\textwidth]{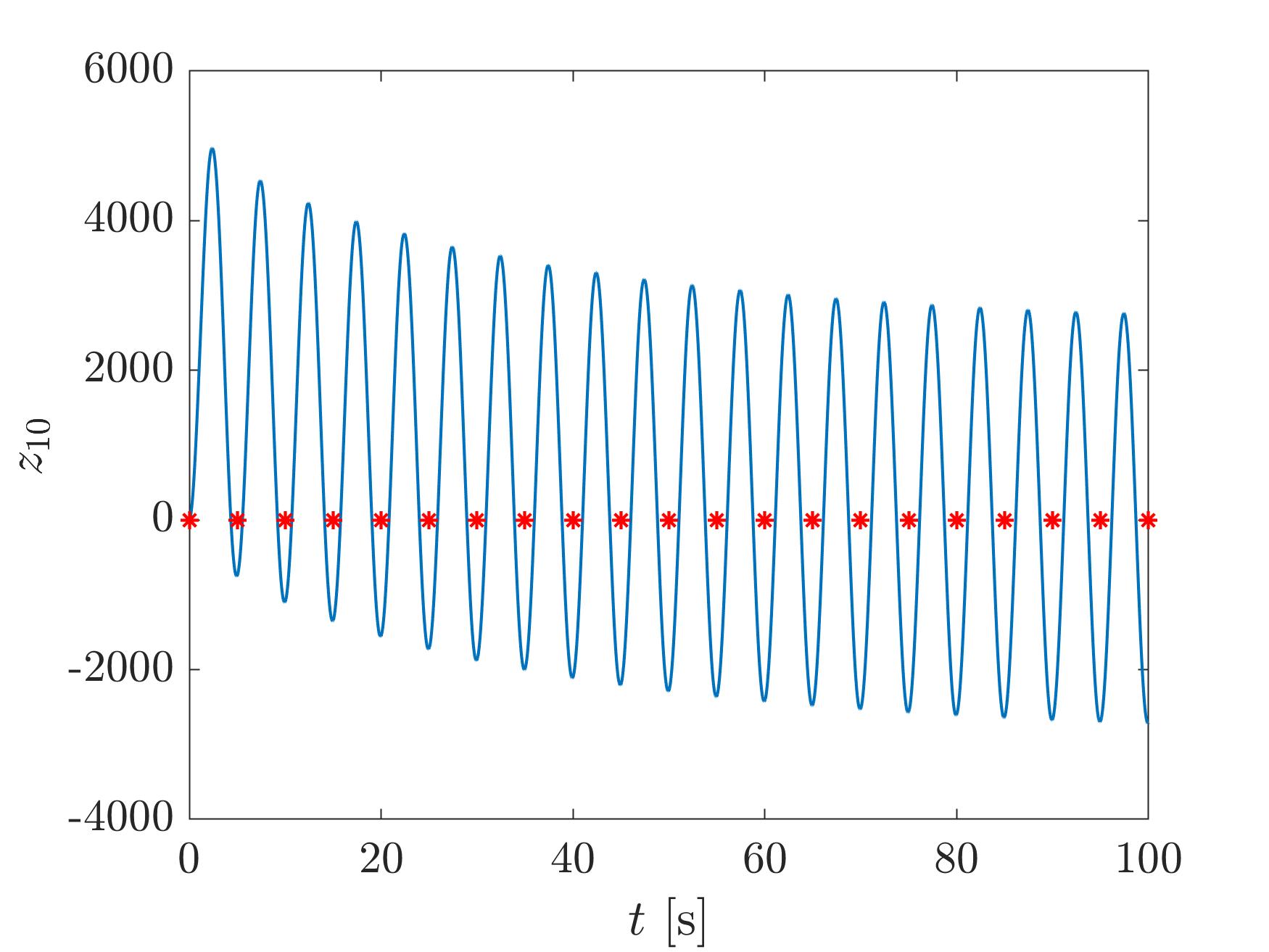}
\caption{$10^{\text{th}}$ internal variable solution using the multi-scale approach.}\label{fig:iv_10_ms}
\end{subfigure}
\caption{}\label{fig:iv_10}
\end{figure}

\begin{figure}[H] 
\centering
\begin{subfigure}{0.45\textwidth}
\includegraphics[width=\textwidth]{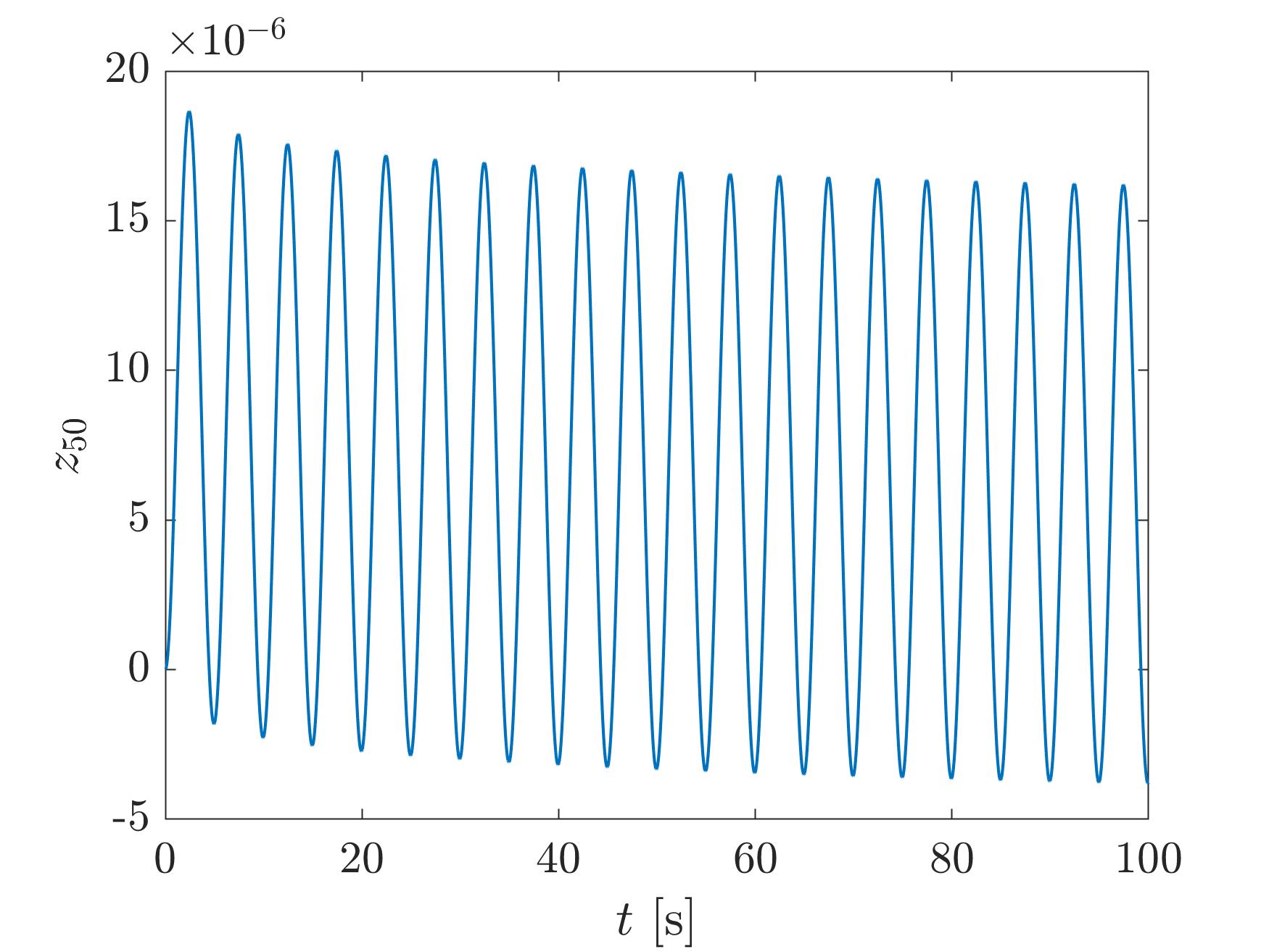}
\caption{$50^{\text{th}}$ internal variable solution using a single-scale approach.}\label{fig:iv_50_classic}
\end{subfigure}
\begin{subfigure}{0.45\textwidth}
\includegraphics[width=\textwidth]{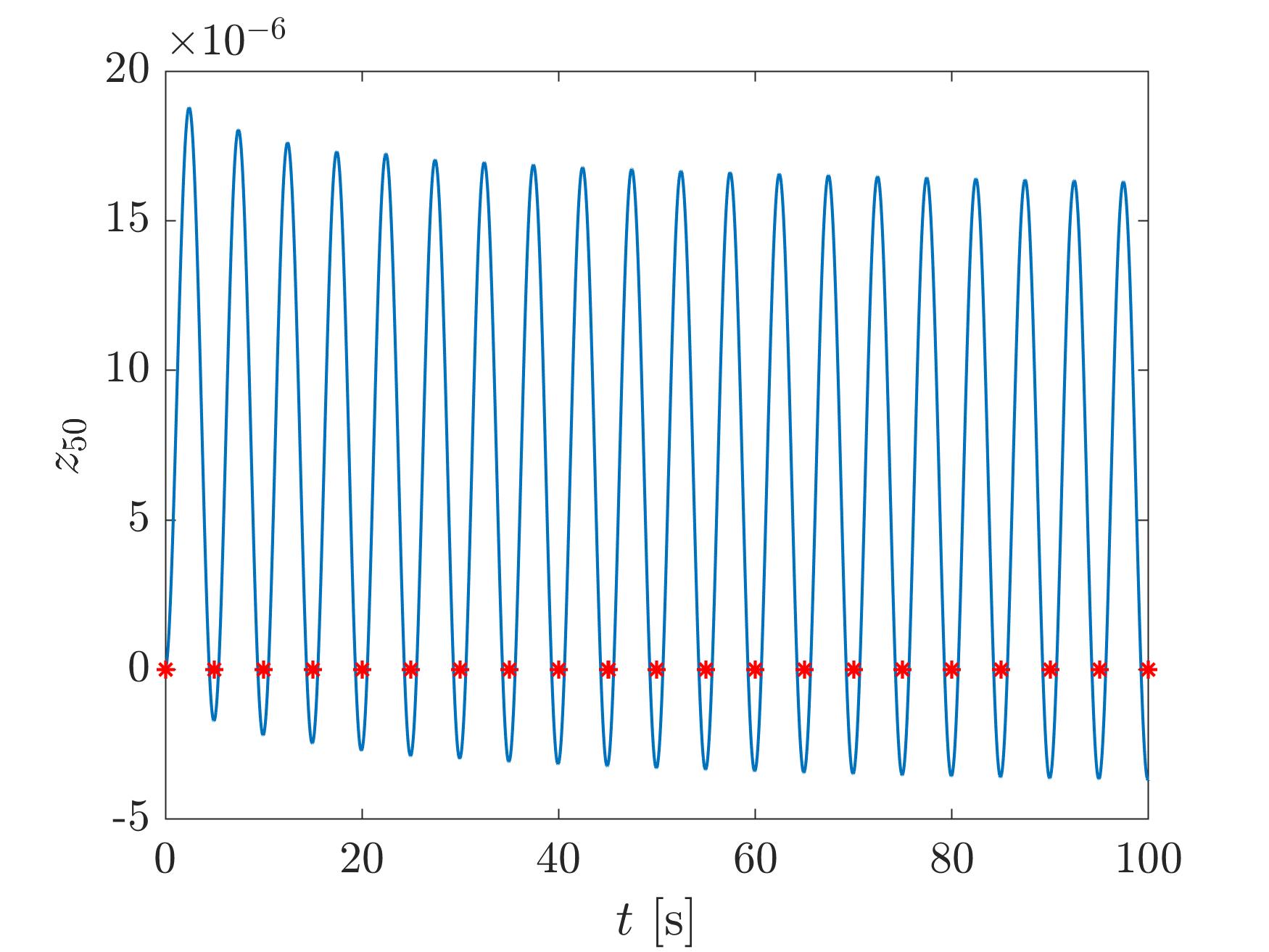}
\caption{$50^{\text{th}}$ internal variable solution using the multi-scale approach.}\label{fig:iv_50_ms}
\end{subfigure}
\caption{}\label{fig:iv_50}
\end{figure}

\begin{figure}[H] 
\centering
\begin{subfigure}{0.45\textwidth}
\includegraphics[width=\textwidth]{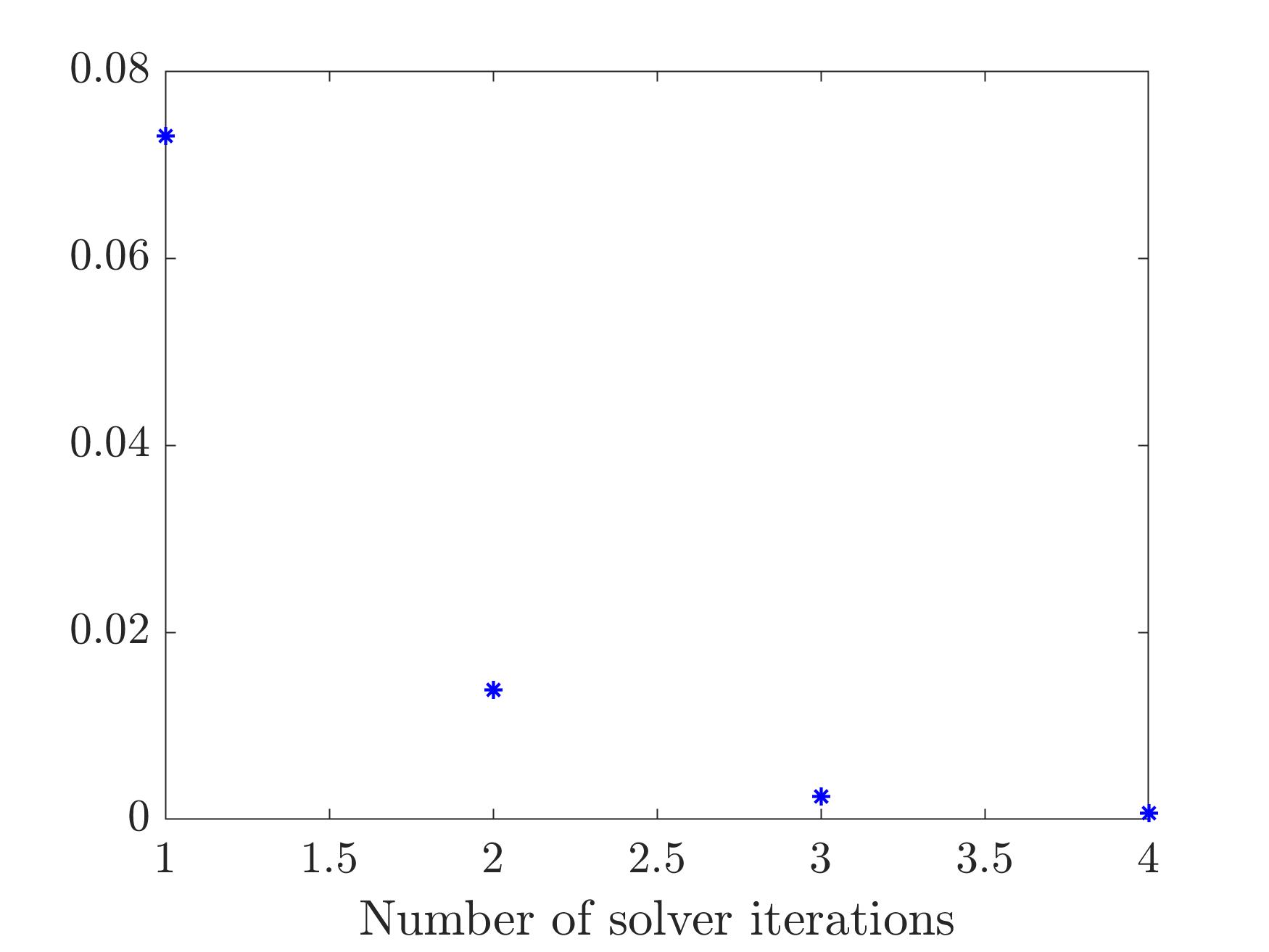}
\caption{Single-scale solution stagnation indicator for $50$ internal variables.}\label{fig:stagnation_50}
\end{subfigure}
\begin{subfigure}{0.45\textwidth}
\includegraphics[width=\textwidth]{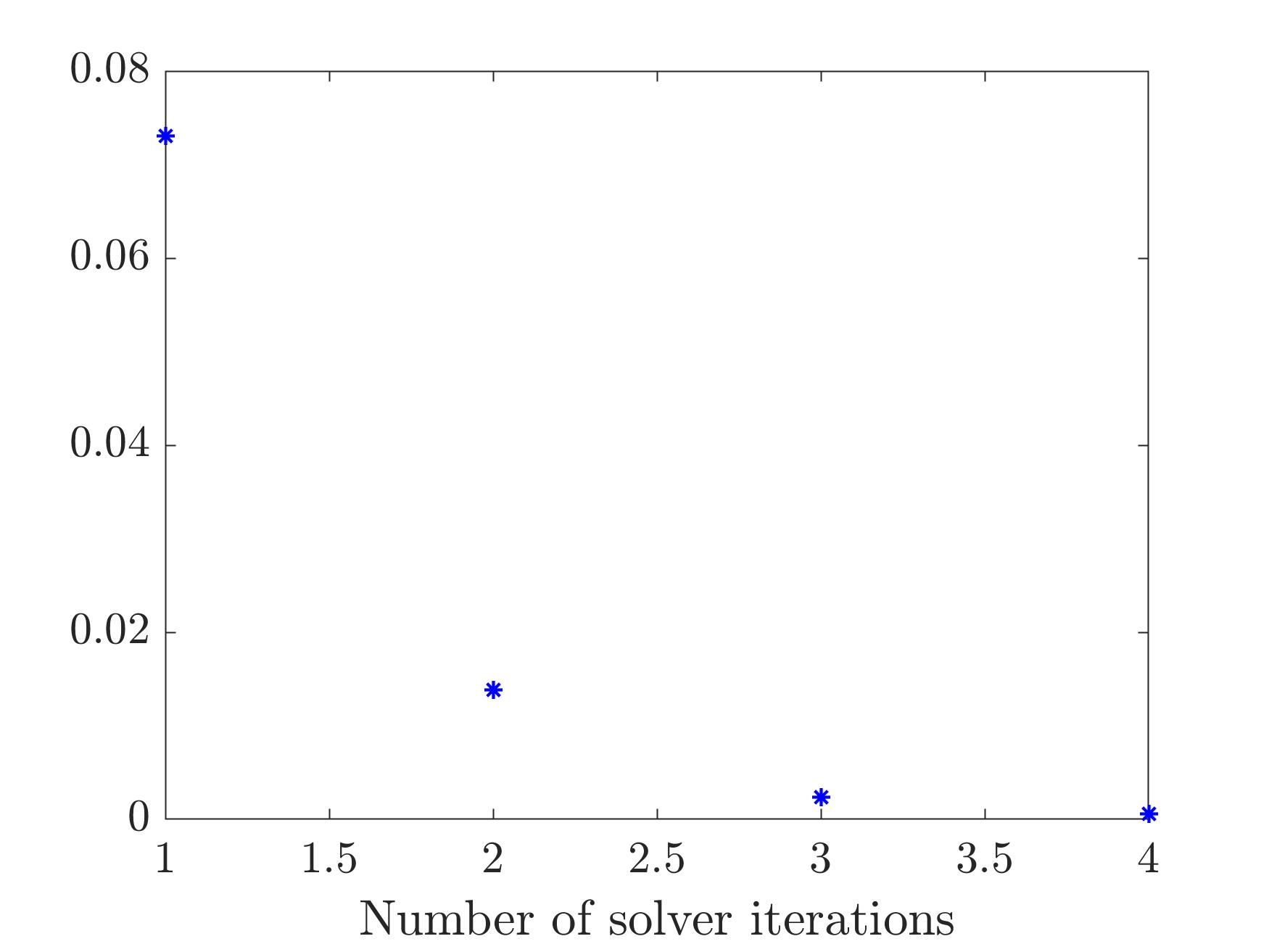}
\caption{Multi-scale solution stagnation indicator for $50$ internal variables.}\label{fig:stagnation_ms_50}
\end{subfigure}
\caption{Stagnation of solvers when solving 50 internal variables.}
\end{figure}

Figures \ref{fig:disp_comp_50} and \ref{fig:disp_comp_ms_50} present the displacement at convergence at the middle of the bar for both single and multi-scale approach.
Figures \ref{fig:iv_10} and \ref{fig:iv_50} present a comparison for the $10^{\text{th}}$ and $50^{\text{th}}$ internal variables at convergence for both single and multi-scale approach.
In addition, figures \ref{fig:stagnation_50} and \ref{fig:stagnation_ms_50} illustrate the curves of stagnation versus number of solver iterations for the single-scale PGD resolution and the one employing the multi-scale approach.
As can be seen from those results, the multi-scale approximation using the Partition of Unity method allows obtaining responses that correctly approximate the reference solution of the system. However, in all the results presented in the present work, the solution using the single-scale PGD was more efficient than its multi-scale approximation. This is due to the costs related to the construction of the necessary operators, as well as to the computation of the multi-scale temporal sub-modes. 

This observation can be explained by the fact that the time interval studied is not sufficiently large. That responds to the fact that the present work aims to present ideas rather than to apply the method to a costly industrial case. Consequently, it is not possible to take full advantage of the multi-scale approximation in a small time interval, however, it is considered as a perspective to study problems where the time interval is very large. In this context, the approach proposed in the present work could reduce the computational costs for its resolution.


\section{Conclusions and perspectives}\label{sec:concl_persp}

In this paper, it has been shown that it is possible to predict viscoelatic behaviors described by internal variables under cyclic loading with the PGD method by considering a globalization of the local models. In addition, a time multi-scale approximation of the temporal functions of the PGD were constructed based on the Partition of Unity method. This choice, allows to construct at a low-cost, the temporal response of the system by exploiting the multi-scale behavior of the system faced to a fatigue load. Something that is of capital importance as the studied model is strongly coupled and leads to high costs when solving large temporal domains.

To highlight the potentiality of the method, the reference problem of section \ref{sec:viscel_problem} was solved using a single-scale PGD and the new proposed multi-scale PGD in time using the PU method when dealing with fatigue load. This work shows that the
PGD combined with a multi-scale approximation in time could be efficiently used to
predict viscoelastic behaviors combined with internal variables under very high cyclic fatigue by decreasing the computational complexity.

As perspectives, we seek to extend this procedure to 3D cases and to include the test of this method with more complex behaviors like nonlinear viscoelasticity. In addition, richer shape functions for the macro behavior approximation should be used in order to decrease the number of modes required of the temporal multi-scale approximation.

\bmhead{Acknowledgements}

Authors acknowledge the support of the ESI Group through its research chair CREATE-ID at Arts et M\'etiers ParisTech.

\section*{Declarations}


\begin{itemize}
\item Funding: This research is part of the program DesCartes and is supported by the National Research Foundation, Prime Minister's Office, Singapore under its Campus for Research Excellence and Technological Enterprise (CREATE) program.
\item Conflict of interest/Competing interests: The authors declare that they have no competing interests.
\item Ethics approval and consent to participate: Yes
\item Consent for publication: Yes
\item Data availability: Provided under request.
\item Materials availability: Not applicable.
\item Code availability: Provided under request.
\item Author contribution: All the authors participated in the definition of techniques and algorithms.
\end{itemize}




\bibliography{sn-article}

\end{document}